\documentclass{iopart}
\usepackage{bbm,amssymb,amsbsy,times,pifont}
\usepackage[dvips]{graphicx}
\usepackage{hyperref}
\usepackage{color}

\newcommand{\prl}{{\it Phys. Rev. Lett.} }
\newcommand{\pra}{{\it Phys. Rev. A} }
\newcommand{\njp}{{\it New J. Phys.} }
\newcommand{\qph}{{\it Preprint} quant-ph/}

\newcommand{\arcsinh}{{\rm arcsinh}}

\newcommand{\id}{\mathbbm{1}}

\renewcommand{\tr}{{\rm Tr}\,}
\renewcommand{\det}{{\rm Det}\,}
\newcommand{\gr}[1]{\boldsymbol{#1}}
\newcommand{\be}{\begin{equation}}
\newcommand{\ee}{\end{equation}}
\newcommand{\bea}{\begin{eqnarray}}
\newcommand{\eea}{\end{eqnarray}}
\newcommand{\ket}[1]{|#1\rangle}
\newcommand{\bra}[1]{\langle#1|}

\newcommand{\N}{{\cal N}}

\newcommand{\sig}{\gr{\sigma}}
\newcommand{\eps}{\gr{\varepsilon}}

\newcommand{\alp}{\gr{\alpha}}

\newcommand{\eq}[1]{Eq.~(\ref{#1})}
\newcommand{\ineq}[1]{Ineq.~(\ref{#1})}
\newcommand{\eg}{\emph{e.g.}~}
\newcommand{\ie}{\emph{i.e.}~}

\begin{document}

\title{\textsf{Optical state engineering, quantum communication,
and robustness of entanglement promiscuity in three-mode Gaussian
states}}

\author{\textsf{Gerardo Adesso$^{1,2,3}$, Alessio Serafini$^{4,5}$ and
Fabrizio Illuminati$^1$}}

\address{$^1$Dipartimento di Fisica ``E. R. Caianiello'',
Universit\`a degli Studi di Salerno; CNR-Coherentia, Gruppo di
Salerno; and INFN Sezione di Napoli-Gruppo Collegato di Salerno; Via
S. Allende, 84081 Baronissi (SA), Italy \\
$^2$Centre for Quantum Computation, DAMTP, Centre for Mathematical
Sciences, University of Cambridge, Wilberforce Road, Cambridge CB3
0WA, United Kingdom \\
$^3$Dipartimento di Fisica dell'Universit\`a ``La Sapienza'' and
Consorzio Nazionale Interuniversitario per le Scienze Fisiche della
Materia, Roma, 00185 Italy \\
$^4$Institute for Mathematical Sciences, Imperial College London, 53
Prince's Gate, SW7 2PE, United Kingdom; and QOLS, The Blackett
Laboratory, Imperial
College London, Prince Consort Road, SW7 2BW, United Kingdom \\
$^5$Department of Physics \& Astronomy, University College London,
Gower Street, London WC1E 6BT, United Kingdom \\
E-mail: \textcolor[rgb]{0,0,1}{gerardo@sa.infn.it},
\textcolor[rgb]{0,0,1}{serale@imperial.ac.uk},
\textcolor[rgb]{0,0,1}{illuminati@sa.infn.it}}

\begin{abstract}
We present a novel, detailed study on the usefulness of three-mode
Gaussian states states for realistic processing of
continuous-variable quantum information, with a particular emphasis
on the possibilities opened up by their genuine tripartite
entanglement. We describe practical schemes to engineer several
classes of pure and mixed three-mode states that stand out for their
informational and/or entanglement properties. In particular, we
introduce a simple procedure -- based on passive optical elements --
to produce pure three-mode Gaussian states with {\em arbitrary}
entanglement structure (upon availability of an initial two-mode
squeezed state). We analyze in depth the properties of distributed
entanglement and the origin of its sharing structure, showing that
the promiscuity of entanglement sharing is a feature peculiar to
symmetric Gaussian states that survives even in the presence of
significant degrees of mixedness and decoherence. Next, we discuss
the suitability of the considered tripartite entangled states to the
implementation of quantum information and communication protocols
with continuous variables. This will lead to a feasible experimental
proposal to test the promiscuous sharing of continuous-variable
tripartite entanglement, in terms of the optimal fidelity of
teleportation networks with Gaussian resources. We finally focus on
the  application of three-mode states to symmetric and asymmetric
telecloning, and single out the structural properties of the optimal
Gaussian resources for the latter protocol in different settings.
Our analysis aims to lay the basis for a practical quantum
communication with continuous variables beyond the bipartite
scenario.
\end{abstract}


\tableofcontents   \title[Optical state engineering, quantum
communication, \ldots in three-mode Gaussian states] \maketitle

\section{Introduction}
The study of {\em multipartite} entanglement in Gaussian states of
continuous variable systems has lately received much attention, both
in view of their interest as a theoretical testground and because of
their versatility towards the effective implementation of
communication protocols. Nevertheless, a complete understanding of
the general features of multipartite quantum correlations and of
their (often controversial) operational interpretation still appears
to be far from accomplished.

For the basic instance of three-mode Gaussian states, a qualitative classification of
multipartite entanglement has been introduced \cite{kraus}, while more recently
a consistent way to quantify such a multipartite entanglement has been presented, with the definition
of the ``residual (Gaussian) contangle'' \cite{3mpra}.
It has also been shown that pure, symmetric three-mode Gaussian states
exhibit a {\em promiscuous} sharing of quantum correlations (where bipartite and genuine multipartite
entanglement are not mutually exclusive but rather reciprocally enhanced) \cite{contangle,3mpra}.
However, several important aspects related to the
complex sharing structure of quantum correlations between the three parties await for further
inspection and clarification.

In this respect, the purpose of this paper is threefold. Firstly, it
investigates the origin of the promiscuous entanglement sharing and
shows that it is crucially related to the global symmetry of the
states (as its signature is still present in mixed, symmetric, while
it is lost in pure, asymmetric states). Secondly, it presents
practical strategies to engineer three-mode Gaussian entangled
states, which may have a remarkable experimental impact towards the
practical realization of optimal resources for specific multipartite
communication tasks. Thirdly, it aims at providing the residual
contangle and its properties with an operational interpretation by
addressing its relationship with the figures of merits of optimized
communication protocols (teleportation networks and telecloning). As
we will see, these two objectives are intimately intertwined, as
they both concern the characterization of the structure of
multipartite entanglement and of its sharing properties.

The article opens with a brief review of the basic properties of
three-mode Gaussian states and of their entanglement
(Sec.~\ref{struct}). We then proceed to investigate the sharing of
quantum correlations in asymmetric, mixed three-mode states,
introducing a paradigmatic class of states (which will be dubbed
``basset hound'' states) and showing that symmetric mixed states
still feature a promiscuous entanglement sharing, whereas pure
asymmetric states lose such a peculiarity (Sec.~\ref{promo}). In
Sec.~\ref{engi}, we describe a novel, `economical' (in a
well-defined sense) strategy to build pure three-mode Gaussian
states with any, arbitrary entanglement structure; furthermore, we
present guidelines for the optical generation of classes of pure and
noisy three-mode Gaussian with relevant entanglement properties.
Sec.~\ref{intrappli} introduces the second part of the paper,
devoted to communication protocols. In Sec.~\ref{secpoppy} we review
and discuss the equivalence between the optimal fidelities of
 three-party teleportation networks and Gaussian residual
contangle in symmetric pure resources, proposing a direct
experimental test of the promiscuous sharing; the demise of the
optimal fidelity under thermal decoherence of the three-mode
resource is also exactly studied and pure symmetric states are shown
to be the most robust (in the specific sense of optimally preserving
the maximal fidelity under decoherence). In Sec.~\ref{sectlc} we
focus on symmetric and asymmetric telecloning, showing that the
operational interpretation of entanglement measures in terms of
teleportation fidelity breaks down for asymmetric states and
determining several instances of three-mode states acting as optimal
(under various constraints) resources for such communication
protocols. Sec.~\ref{conclu} concludes the paper, finalizing the
line of work undertaken in Ref.~\cite{3mpra} and completed in the
present article.

\section{Three-mode Gaussian states: structural and entanglement properties}\label{struct}
We consider a continuous variable (CV) system consisting of $N$
bosonic modes, associated to an infinite-dimensional Hilbert space
${\cal H}$ and described by the vector $\hat{X}=\{\hat x_1,\hat
p_1,\ldots,\hat x_N,\hat p_N\}$ of the field quadrature operators,
whose canonical commutation relations can be expressed in matrix
form: $[\hat X_{i},\hat X_j]=2i\Omega_{ij}$, with the symplectic
form $\Omega=\oplus_{i=1}^{n}\omega$ and $\omega=\delta_{ij-1}-
\delta_{ij+1},\, i,j=1,2$.

Quantum states of paramount importance in CV systems are the
so-called Gaussian states, {\em i.e.}~states with Gaussian
characteristic functions and quasi-probability distributions
\cite{CVbook1,CVbook2,review,mytesi}. The interest in this special
class of states (which includes vacua, coherent, squeezed, thermal,
and squeezed-thermal states of the electromagnetic field) stems from
the feasibility to produce and control them with linear optical
elements, and from the increasing number of efficient proposals and
successful implementations of quantum information and communication
protocols involving multimode Gaussian states. For a review of the
basic properties of Gaussian states and their structural and
entanglement characterization, see {\em e.g.}~Ref.~\cite{mytesi}. A
more concise (and closely related to the context of this paper)
 background is provided in Ref.~\cite{3mpra}, which is
focused on three-mode states and contains a quite general
introduction to phase-space formalism and symplectic operations
(making use of the same notation adopted here).  In this section, we
limit ourselves to define the relevant notation and recall some
useful results.

Neglecting first moments (which can be arbitrarily adjusted by local
unitaries), one can completely characterize a Gaussian state by the
real, symmetric covariance matrix (CM) $\gr{\sigma}$, whose entries
are $\sigma_{ij}=1/2\langle\{\hat{X}_i,\hat{X}_j\}\rangle
-\langle\hat{X}_i\rangle\langle\hat{X}_j\rangle$. Throughout the
paper
 $\gr{\sigma}$ will be used indifferently to indicate the CM
of a Gaussian state or the state itself. The CM $\gr{\sigma}$ must
fulfill the Robertson-Schr\"odinger uncertainty relation
$\gr{\sigma}+i\Omega \geq 0$ \cite{simon87}.

For future convenience, let us write down the CM  $\sig$ of a
$3$-mode Gaussian state in terms of two by two submatrices as \be
\sig = \left(\begin{array}{ccc}
\sig_{1} & \eps_{12} & \eps_{13} \\
\eps_{12}^{\sf T} & \sig_{2} & \eps_{23} \\
\eps_{13}^{\sf T} & \eps_{23}^{\sf T} & \sig_{3} \\
\end{array}\right) \; . \label{subma}
\ee
In what follows, we will refer to three-mode Gaussian states endowed
with symmetries under mode exchange as ``bisymmetric'' (invariant
under the permutation of a specific pair of modes) and ``fully
symmetric'' states (invariant under the permutation of any pair of
modes). In terms of the $2\times 2$ blocks of \eq{subma}, the CM of three-mode
bisymmetric states is defined by $\sig_j=\sig_k$ and $\eps_{jl}=\eps_{kl}$
for indexes $j\neq k\neq l$ with values between $1$ and $3$.
Clearly, fully symmetric states have
\be
\gr\alpha \equiv \sig_1= \sig_2 =\sig_3 \quad {\rm and}
\quad \gr\zeta \equiv \eps_{12}=\eps_{13}=\eps_{23} \, .  \label{fscm}
\ee
The entanglement of bisymmetric
states can be concentrated in two modes by local unitary operations
\cite{unitarily,adescaling}.\footnote{A pictorial description of
such unitary-localization procedure is provided in the following
(see Fig. \ref{figbasset}).} In general, for $(M+N)$-mode states,
bisymmetric states are defined as invariant under the exchange of any two modes within
the subsystems of $M$ and $N$ modes.

\subsection{Separability properties}\label{sepa2m}

The positivity of the partially transposed CM $\tilde{\gr{\sigma}}$
has been proven to be necessary and sufficient for the separability
of $(1+N)$-mode and bisymmetric $(M+N)$-mode Gaussian states
\cite{simon00,duan00,wernerwolf,unitarily}, providing a clearcut
qualitative characterization of the entanglement of such states (PPT
criterion). An ensuing computable measure of CV entanglement is the
\emph{logarithmic negativity} \cite{vidwer} $E_{\N}\equiv
\ln\|\tilde{\varrho}\|_{1}$, where $\| \cdot \|_1$ denotes the trace
norm, which constitutes an upper bound to the {\em distillable
entanglement} of the state $\varrho$. For Gaussian states, it can be
computed in terms of the symplectic spectrum $\tilde{\nu}_i$ of the
partially transposed CM $\tilde{\gr{\sigma}}$ \cite{3mpra}:
\begin{equation}\label{ensy}
E_{\N}= \max\left\{0,\,-{\sum}_{i :
\tilde{\nu}_i<1}\ln\tilde{\nu}_i\right\}\,.
\end{equation}

A complete {\em qualitative} characterization of the entanglement of
three-mode Gaussian state has been obtained in \cite{kraus}, where
five different instances have been classified according to their separability
properties under different bipartitions.
Let us also recall that, from a more practically-oriented viewpoint,
efficient criteria to detect multipartite entanglement from the
knowledge of the second moments
have been developed in \cite{vanlokfuru}.
Moreover, another effective way of testing the
presence of bipartite and/or genuine tripartite entanglement in
Gaussian states is through {\em entanglement witnesses}, based on
linear and nonlinear functionals of the CM \cite{illuso}.

As for the {\em quantification} of the tripartite entanglement of
Gaussian states, it has been shown in \cite{contangle,hiroshima}
that, assuming the {\em `contangle'} $E_{\tau}^{j|k}$ as an
entanglement measure (formally defined as the convex roof of the
squared logarithmic negativity, here the notation $j|k$ refers to
some pair of subsystems), a `monogamy' inequality \cite{sharing}
analogous to the finite dimensional one \cite{CKW,Osbornio} holds.
For three modes, the inequality reads:
$E_{\tau}^{j|(kl)}-E_{\tau}^{j|k}-E_{\tau}^{j|l}\ge 0$, where
$j,k,l$ label the three modes and can be permuted at will. One can
also define the {\em Gaussian contangle} $G_{\tau}^{j|k}$, where the
convex roof is restricted to Gaussian decompositions. Clearly, the
Gaussian contangle is in general easier to handle analytically than
the contangle; it fulfills the monogamy inequality as well. The
monogamy constraints suggest proper quantifiers of genuine
tripartite entanglement. In particular, the minimum {\em residual
contangle} $E_\tau^{res}$ \cite{contangle} is defined as
\begin{equation}\label{etaumin}
E_\tau^{res}\equiv\min_{(i,j,k)} \left[
E_\tau^{i|(jk)}-E_\tau^{i|j}-E_\tau^{i|k}\right]\,,
\end{equation}
where $(i,j,k)$ denotes all the permutations of the three mode
indexes. Likewise, the minimum residual
Gaussian contangle $G_\tau^{res}$, compactly referred to as {\em
arravogliament} (or ``arravojament'') \cite{contangle,3mpra}, reads
\begin{equation}\label{gtaures}
G_\tau^{res} \equiv G_\tau^{i|j|k}\equiv\min_{(i,j,k)} \left[
G_\tau^{i|(jk)}-G_\tau^{i|j}-G_\tau^{i|k}\right]\,.
\end{equation}
Such a measure is monotonic under Gaussian local operations and classical communication
\cite{contangle} and will be adopted to quantify the genuine tripartite entanglement of
three-mode Gaussian states. The operational status of this measure will be discussed in the
final part of the paper, in terms of teleportation fidelities.

As demonstrated in Ref. \cite{3mpra}, the three local symplectic
invariants $\det{\sig_{1}}$, $\det{\sig_{2}}$ and $\det{\sig_{3}}$
(strictly related to the local mixednesses of the state
\cite{parisera}) fully determine the entanglement of any possible
bipartition of a pure three-mode Gaussian state with CM $\sig$, and
also the genuine tripartite entanglement contained in the state
\cite{contangle}. Moreover, defining $a_j \equiv
\sqrt{\det{\sig_j}}$ for $j=1,\ldots,3$ one has that the three
single-mode mixednesses are constrained by the following `triangle'
entropic inequality (see \cite{3mpra} for a detailed proof) \be
\label{triangleprim} |a_i-a_j|+1\le a_k \le a_i+a_j-1 \; . \ee
Remarkably, Inequality (\ref{triangleprim}) (together with the
conditions $a_l \ge 0\,\,\forall l$) fully characterizes the local
symplectic eigenvalues of the CM of three-mode pure Gaussian states,
thus providing a complete characterization of the entanglement of
such states. In fact, the CM of \eq{subma} can be put in a {\em
standard form} completely parametrized by the three local
mixednesses, with $\sig_i = {\rm diag}\{a_i,\,a_i\}$ and the blocks
$\eps_{ij}$ diagonal as well, with elements only functions of the
$\{a_i\}$'s \cite{3mpra}.

\section{Promiscuous entanglement sharing versus noise and asymmetry}\label{promo}

In Ref.~\cite{contangle}, as a direct application of the monogamy
constraint on the distribution of tripartite quantum correlations,
the entanglement sharing structure of three-mode Gaussian states has
been investigated. In particular, it has been demonstrated that
pure, fully symmetric (\ie with $\det\sig_1=\det\sig_2=\det\sig_3
\equiv a$ in \eq{subma}) Gaussian states exist which are maximally
three-way entangled and, at the same time, possess the maximum
possible entanglement between any pair of modes in the corresponding
two-mode reduced states. Those states, known as CV finite-squeezing
GHZ/$W$ states, are thus said to exhibit a {\em promiscuous}
entanglement sharing \cite{contangle,3mpra}. The notion of
``promiscuity'' basically means that bipartite and genuine
multipartite (in this case tripartite) entanglement are increasing
functions of each other, and the genuine tripartite entanglement is
enhanced by the simultaneous presence of the bipartite one, while
typically in low-dimensional systems like qubits only the opposite
behaviour is compatible with monogamy \cite{CKW,wstates}. The
promiscuity of entanglement in three-mode GHZ/$W$ states is,
however, {\em partial}\footnote{It can be in fact shown that in
Gaussian states of CV systems with more than three modes,
entanglement can be distributed in an {\em infinitely} promiscuous
way \cite{unlim}.}. Namely they exhibit, with increasing squeezing,
an  unlimited tripartite entanglement given by \cite{3mpra}
\begin{equation}\label{gresghzw}
\hspace*{-1.3cm}G_\tau^{res}(\sig^{_{{\rm GHZ}/W}})=
\arcsinh^2\!\left[\sqrt{a^2 -1}\right] - \frac{1}{2}
\ln^2\!\left[\frac{3 a^2 - 1 -\sqrt{9 a^4 - 10 a^2 +
1}}{2}\right]\,,
\end{equation}
and thus  diverging in the limit $a \rightarrow \infty$. At the same
time, they also possess a nonzero,  accordingly increasing bipartite
entanglement between any two modes, which nevertheless stays finite
even for infinite squeezing. Precisely, from \eq{gresghzw}, it
saturates to the value
\begin{equation}\label{gredmaxghzw}
G_\tau^{i|j}(\sig^{_{{\rm GHZ}/W}},\,a \rightarrow \infty) =
\frac{\ln^2{3}}{4} \approx 0.3\,.
\end{equation}

We will see later on how this promiscuous distribution of
entanglement can be demonstrated experimentally in terms of CV
teleportation experiments. Prior to that, it is natural to question
whether {\em all} three-mode Gaussian states  are expected to
exhibit a promiscuous entanglement sharing.
So far, such a question has not yet been addressed.
We shall therefore investigate the robustness of promiscuity against the lack of each of the two
defining properties of GHZ/$W$ states: full symmetry, and global
purity. We will specifically find that promiscuity survives under
mixedness, but is in general lost if the complete
permutation-invariance is relaxed.

\subsection{Promiscuity versus mixedness: Entanglement distribution in noisy GHZ/{\em W} states}
\label{secnoisyghzw}

We consider here the noisy version of the three-mode GHZ/$W$ states,
which are a family of mixed Gaussian fully symmetric states, also
called three-mode squeezed thermal states \cite{3mcinese}. They
result in general from the dissipative evolution of pure GHZ/$W$
states in proper Gaussian noisy channels \cite{3mpra}.
 Let us mention
that various properties of noisy three-mode Gaussian states have
already been addressed, mainly regarding their effectiveness in the
implementation of CV protocols \cite{pirandolo,paris05}. However, here we
focus on the properties of genuine multipartite entanglement of such states,
which have not been considered previously.
This analysis will allow us to go beyond the set of pure states,
thus gaining deeper insight into the role played by realistic
quantum noise in the sharing and characterization of tripartite
entanglement.

Noisy GHZ/$W$ states are described by a fully symmetric CM $\sig_s^{th}$ of the form
(\ref{fscm}), with $\gr\alpha=a \id_2$ and $\gr\zeta={\rm
diag}\{e^+,\,e^-\}$, where
\begin{equation}\label{epmthermal}
e^\pm = \frac{a^2 - n^2 \pm \sqrt{\left(a^2 - n^2\right) \left(9 a^2
- n^2\right)}}{4 a}\,,
\end{equation}
where $a \ge n$ to ensure the physicality of the state. These states
are thus completely determined by the local purity $\mu_l=a^{-1}$
and by the global purity $\mu = n^{-3}$.\footnote{The ``purity''
$\mu$ fo a quantum state $\varrho$ is given by $\tr{(\varrho^2)}$.
For a Gaussian state $\varrho$ with CM $\sig$ one has
$\mu=1/\sqrt{\det\sig}$ \cite{parisera}.} For ease of notation, let
us replace the parameter $a$ with the ``'squeezing parameter'' $s$,
defined by $3as=n\sqrt{2s^4+5s^2+2}$ (whose physical significance
will become clear once the experimental generation of GHZ/$W$ will
be described in Sec.~\ref{SecEngiGHZW}). Noisy GHZ/$W$ states reduce
to pure GHZ/$W$ states ({\ie}three-mode squeezed {\em vacuum}
states) for $n=1$.

\subsubsection{Separability properties} Depending on the defining parameters $s$ and $n$, noisy GHZ/$W$
states can belong to three different separability classes according
to the classification of Ref. \cite{kraus}. Namely, as explicitly
computed in Ref.~\cite{3mcinese}, we have in our notation
\begin{eqnarray}
\hspace*{-1cm}
  s > \frac{\sqrt{9 n^4 - 2 n^2 + 9 +
          3 \left(n^2 - 1\right) \sqrt{9 n^4 + 14 n^2 + 9}}}{4 n}
          &\Rightarrow& \mbox{\it Class 1;} \label{termic1} \\
\hspace*{-1cm}  n < s \le \frac{\sqrt{9 n^4 - 2 n^2 + 9 +
          3 \left(n^2 - 1\right) \sqrt{9 n^4 + 14 n^2 + 9}}}{4 n}
          &\Rightarrow& \mbox{\it Class 4;}  \label{termic4}\\
\hspace*{-1cm}  s \le n &\Rightarrow& \mbox{\it Class
5.}\label{termic5}
\end{eqnarray}
States which fulfill \ineq{termic1} are fully inseparable (Class 1,
encoding genuine tripartite entanglement), while states that violate
it have positive partial transposition with respect to all
bipartitions. However, in this case PPTness does not imply
separability. In fact, in the range defined by \ineq{termic4}, noisy
GHZ/$W$ states are three-mode biseparable (Class 4), that is they
exhibit tripartite {\em bound entanglement}. This can be verified by
showing, using the methods of Ref.~\cite{kraus}, that such states
cannot be written as a convex combination of separable states.
Finally, noisy GHZ/$W$ states that fulfill \ineq{termic5} are fully
separable (Class 5), containing no entanglement at all.

\begin{figure}[t!]
\centering
\includegraphics[width=9cm]{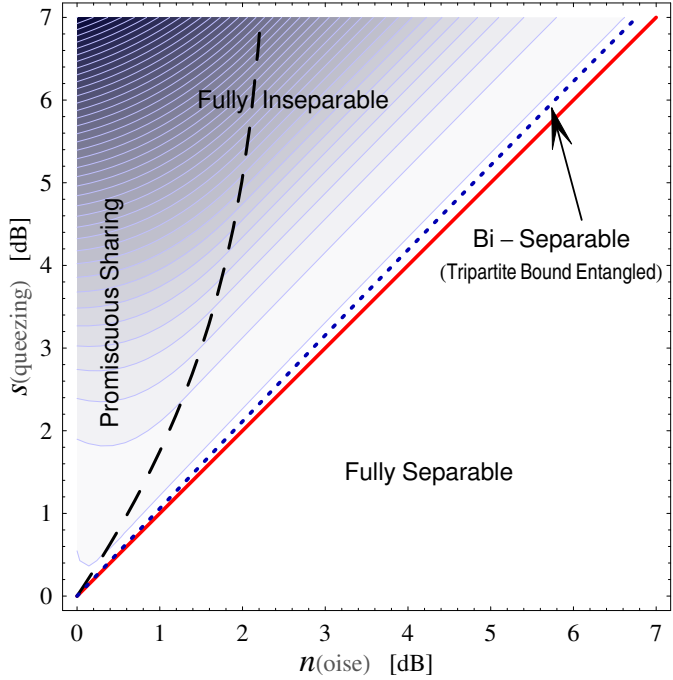}
\caption{Summary of separability and entanglement properties of
three-mode squeezed thermal states, or noisy GHZ/$W$ states, in the
space of the two parameters $n$ and $s$. The separability is
classified according to the scheme of Ref. \cite{kraus}. Above the
dotted line the states are fully inseparable (Class 1); below the
solid line they are fully separable (Class 5). In the narrow
intermediate region, noisy GHZ/$W$ states are three-mode biseparable
(Class 4), {\ie}they exhibit tripartite bound entanglement. The
relations defining the boundaries for the different regions are
given in Eqs.~{(\ref{termic1}--\ref{termic5})}. In the fully
inseparable region, the residual (Gaussian) contangle \eq{gtauresth}
is depicted as a contour plot, growing with increasing darkness from
$G_\tau^{res}=0$ (along the dotted line) to $G_\tau^{res}\approx
1.9$ (at $n=0$ dB, $s=7$ dB). On the left side of the dashed line,
whose expression is given by \eq{termprom}, not only genuine
tripartite entanglement is present, but also each reduced two-mode
bipartition is entangled. In this region, $G_\tau^{res}$ is strictly
larger than in the region where the two-mode reductions are
separable. This evidences the {\em promiscuous} sharing structure of
multipartite CV entanglement in symmetric, even mixed, three-mode
Gaussian states.} \label{figacunt}
\end{figure}

The tripartite residual Gaussian contangle of \eq{gtaures}, which is
nonzero only in the fully inseparable region, can be explicitly
computed. In particular, the $1 \times 2$ Gaussian contangle
$G_\tau^{i|(jk)}$ is obtained by exploiting the {\em unitary
localizability} of entanglement in symmetric Gaussian states
\cite{unitarily}. Namely, if one lets modes $2$ and $3$ interfere at
a 50:50 beam-splitter, this operation (local unitary with respect to
the imposed $1|(23)$ bipartition) decouples the transformed mode
$3'$ (more details will be given later, see Fig.~\ref{figbasset}).
Moreover, one finds that the resulting equivalent two-mode state of
modes $1$ and $2'$ is symmetric, and so the bipartite contangle
between mode $1$ and the block of modes $(23)$, equal to the
bipartite contangle between modes $1$ and $2'$, coincides with the
squared logarithmic negativity \cite{contangle}. As for the two-mode
Gaussian contangles $G_\tau^{1|2}=G_\tau^{1|3}$, the same result
holds, as the reduced states are symmetric. Finally one gets, in the
range defined by \ineq{termic1}, a tripartite entanglement given by
\begin{eqnarray} \label{gtauresth}
  G_\tau^{res}(\sig_s^{th}) &=& \frac14 \ln^2\!\left\{\frac{n^2 \left[4 s^4 + s^2 +4 -
          2 \left(s^2 - 1\right) \sqrt{4 s^4 + 10 s^2 + 4}\right]}{9 s^2} \right\}
  \nonumber \\
   &-& 2 \left[\max \left\{0, -\ln \left(\frac{n \sqrt{s^2 +
                    2}}{\sqrt{3} s}\right)\right\}\right]^2\,,
\end{eqnarray}
and $G_\tau^{res}(\sig_s^{th})=0$ when \ineq{termic1} is violated.
For noisy GHZ/$W$ states, the residual Gaussian contangle
\eq{gtauresth} is still equal to the true one \eq{etaumin} (like in
the special instance of pure GHZ/$W$ states), thanks to the symmetry
of the two-mode reductions, and of the unitarily transformed state
of modes $1$ and $2'$.

\subsubsection{Sharing structure} The second term in \eq{gtauresth} embodies the sum of
the couplewise entanglement in the $1|2$ and $1|3$ reduced
bipartitions. Therefore, if its presence enhances the value of the
tripartite residual contangle (as compared to what happens if it
vanishes), then one can infer that entanglement sharing is
`promiscuous' in the (mixed) three-mode squeezed thermal Gaussian
states as well (`noisy GHZ/$W$' states). And this is exactly the
case, as shown in the contour plot of Fig.~\ref{figacunt}, where the
separability and entanglement properties of noisy GHZ/$W$ states are
summarized, as functions of the parameters $n$ and $s$ expressed in
decibels.\footnote{\label{notedb} The noise expressed in decibels
(dB) is obtained from the covariance matrix elements via the formula
$N_{ij}(dB) = 10 \log_{10}(\sigma_{ij})$.} Explicitly, one finds
that for
\begin{equation}\label{nsoglia}
n \ge \sqrt{3}\,,
\end{equation}
corresponding to $\approx 2.386$ dB of noise, the entanglement
sharing can never be promiscuous, as the reduced two-mode
entanglement is zero for any (even arbitrarily large) squeezing $s$.
Otherwise, applying PPT criterion to any two-mode reduction one
finds that for sufficiently high squeezing $s$ bipartite
entanglement is also present in any two-mode reduction, namely
\begin{eqnarray}
  n<\sqrt{3}\quad {\rm and} \quad s > \frac{\sqrt{2} n}{\sqrt{3-n^2}}
          &\quad\Rightarrow\quad& \mbox{{\em promiscuous} sharing}\,. \label{termprom}
\end{eqnarray}

Evaluation of \eq{gtauresth}, as shown in Fig.~\ref{figacunt},
clearly demonstrates that the genuine tripartite entanglement
increases with increasing bipartite entanglement in any two-mode
reduction, unambiguosly confirming that CV quantum correlations
distribute in a promiscuous way not only in pure
\cite{contangle,3mpra}, but also in {\em mixed} symmetric three-mode
Gaussian states. However, the global mixedness is prone to affect
this sharing structure, which is completely destroyed if, as can be
seen from \eq{nsoglia}, the global purity $\mu$ falls below
$1/(3\sqrt{3}) \approx 0.19245$. This purity threshold is remarkably
low: a really strong amount of global noise is necessary to destroy
the promiscuity of entanglement distribution.

We will now provide an example of three-mode states with weaker symmetry constraints,
where the entanglement exhibits a more traditional sharing structure,
\ie with bipartite and tripartite quantum correlations being mutual competitors.

\subsection{Promiscuity versus lack of symmetry: basset hound states} \label{secbas}

Let us  consider here an instance of tripartite entangled states
which are not fully symmetric, but merely bisymmetric pure Gaussian
states. Bisymmetry (a property definable for $M \times N$ Gaussian
states \cite{unitarily}) means in this case  invariance under the
exchange of modes $2$ and $3$, and not under the exchange of {\em
any} two modes as in the previous (pure and mixed) GHZ/$W$
instances.

\begin{figure}[t!]
\centering{\includegraphics[width=6cm]{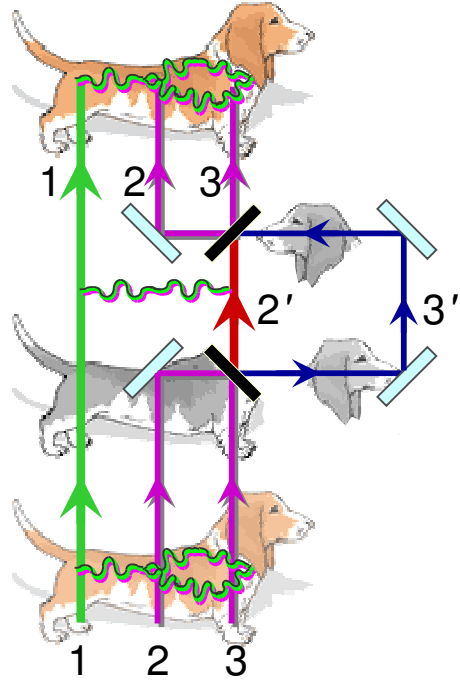}%
\caption{\label{figbasset} {\sc If You Cut The Head Of A Basset
Hound, It Will Grow Again} \cite{notebasset}. Graphical depiction of
the process of unitary localization (concentration) and
delocalization (distribution) of entanglement in three-mode
bisymmetric Gaussian states \cite{unitarily} (or ``basset hound''
states), described in the text. Initially, mode $1$ is entangled
(entanglement is depicted as a waving string) with both modes $2$
and $3$. It exists a local (with respect to the $1|(23)$
bipartition) symplectic operation, realized {\eg}via a beam-splitter
(denoted by a black thick dash), such that all the entanglement is
concentrated between mode $1$ and the transformed mode $2'$, while
the other transformed mode $3'$ decouples from the rest of the
system ({\em unitary localization}). Therefore, the head of the
basset hound (mode $3'$) has been cut off. However, being realized
through a symplectic operation ({\ie}unitary on the density matrix),
the process is reversible: operating on modes $2'$ and $3'$ with the
inverse symplectic transformation, yields the original modes $2$ and
$3$ entangled again with mode $1$, without any loss of quantum
correlations ({\em unitary delocalization}): the head of the basset
hound is back again.}}
\end{figure}

Following the arguments summarized in Fig.~\ref{figbasset},
bisymmetric Gaussian states will be in general referred to as {\em
basset hound} states. To our aims, it is sufficient here to consider
{\em pure} basset hound states\footnote{Unless explicitly stated, we
will always assume in the following that the expression ``basset
hound state'' denotes a {\em pure} bisymmetric Gaussian state.
Notice that the possibility of unitary localization is common to all pure states \cite{holwer,botrez,giedke03}
and is not
specifically related -- {\em for pure states} -- to their symmetry properties.}.
Such states are characterised by a CM $\sig_{B}^p$ of the form \eq{subma},
with
\begin{eqnarray}
  &   \sig_1 = a \id_2,\quad &
  \sig_2 = \sig_3 =  \left(\frac{a+1}{2}\right) \id_2,
  \label{bassigl} \\
  &   \eps_{23} = \left(\frac{a-1}{2}\right) \id_2,\quad &
  \eps_{12} = \eps_{13} = {\rm
  diag}\left\{\frac{\sqrt{a^2-1}}{\sqrt{2}},\,-\frac{\sqrt{a^2-1}}{\sqrt{2}}\right\}.
  \label{basseps}
\end{eqnarray}
They belong to a family of states introduced in
Ref.~\cite{telecloning} as resources for optimal CV telecloning
({\ie}cloning at distance, or equivalently teleportation to more
than one receiver) of single-mode coherent states. We shall
discuss this protocol in detail in Sec.~\ref{sectlc}.

\subsubsection{Tripartite entanglement}
 From a qualitative point of view, basset hound states are fully
inseparable for $a>1$ and fully separable for $a=1$, as already
remarked in Ref.~\cite{telecloning}; moreover, the PPT criterion
 entails that the two-mode reduced state of
modes $2$ and $3$ is always separable. Following the guidelines of
Sec.~\ref{sepa2m}, the residual Gaussian contangle $G_\tau^{res}$
of such states is easily computable.
As we know \cite{contangle}, the
minimum in \eq{gtaures} is attained by choosing as probe the mode of
smallest local mixedness. In our setting, this corresponds to set
either mode $2$ or mode $3$ (indifferently, due to the bisymmetry)
to be the probe mode. Let us choose mode $3$; then we have
\begin{equation}\label{gtauresbh}
G_\tau^{res}(\sig_B^p) = G_\tau^{3|(12)}(\sig_B^p) -
G_\tau^{3|1}(\sig_B^p)\,,
\end{equation}
with
\begin{eqnarray}
G_\tau^{3|(12)}(\sig_B^p) &=& \arcsinh^2\left[\frac{1}{2}
\sqrt{(a - 1) (a + 3)}\right]\,, \label{g12bh}\\
G_\tau^{3|1}(\sig_B^p) &=& \arcsinh^2 \left[\sqrt{\frac{(3 a + 1)^2}
{(a + 3)^2} - 1}\right]\,. \label{g11bh}
\end{eqnarray}
The tripartite entanglement of \eq{gtauresbh} is strictly smaller than
that of GHZ/$W$ states, but it can still diverge in the limit of
infinite squeezing ($a \rightarrow \infty$) due to the global purity
of these basset hound states. Instead, the bipartite entanglement
$G_\tau^{1|2}=G_\tau^{1|3}$ between mode $1$ and each of the modes
$2$ and $3$ in the corresponding two-mode reductions, given by
\eq{g11bh}, is strictly {\em larger} than the bipartite entanglement
in any two-mode reduction of GHZ/$W$ states. This does not
contradict the  previously given characterization of GHZ/$W$ states
as maximally three-way and two-way entangled (maximally
promiscuous). In fact, GHZ/$W$ states have maximal couplewise
entanglement between {\em any} two-mode reduction, while in basset
hound states only two (out of three) two-mode reductions are
entangled, allowing this entanglement to be larger. This is the
reason why these states are well-suited for telecloning, as we will
detail in Sec. \ref{telebass}. Nevertheless, this reduced bipartite
entanglement cannot increase arbitrarily in the limit of infinite
squeezing, because of the monogamy inequality satisfied by the Gaussian contangle.
In fact, it saturates to
\begin{equation}\label{gredmaxbass}
G_\tau^{1|l}(\sig_B^p,\,a \rightarrow \infty) = \ln ^2\left[3 + 2
\sqrt{2}\right] \approx 3.1\,,
\end{equation}
which is about ten times the asymptotic value of the reduced
bipartite two-mode entanglement for GHZ/$W$ states,
\eq{gredmaxghzw}.

\subsubsection{Sharing structure} It is interesting to notice that entanglement
sharing in basset hound states is {\em not} promiscuous.  Tripartite
and bipartite entanglement coexist (the latter only in two of the
three possible two-mode reductions), but the presence of a strong
bipartite entanglement does not help the tripartite one to be
stronger (at fixed local mixedness $a$) than in other states, like
GHZ/$W$ states or even $T$ states \cite{contangle} (which are
globally mixed and moreover contain no reduced bipartite
entanglement at all).

\subsection{The origin of tripartite entanglement promiscuity?}

The above analysis of the entanglement sharing structure in
three-mode Gaussian states (including the non-fully-symmetric basset
hound states, whose entanglement structure is not promiscuous)
delivers a clear hint that, in the tripartite Gaussian setting, {\em
`promiscuity'} is a peculiar consequence not of the global purity
(noisy GHZ/$W$ states remain promiscuous for quite strong
mixedness), but of the complete {\em symmetry} under modes-exchange.
Beside frustrating the maximal entanglement between pairs of modes
\cite{frusta}, symmetry also constrains the multipartite sharing of
quantum correlations. In basset hound states (bisymmetric Gaussian
states), the separability of the reduced state of modes $2$ and $3$
prevents the three modes from having a strong genuine tripartite
entanglement among them all, despite the heavy quantum correlations
shared by the two couples of modes $1|2$ and $1|3$.

It is instructive to recall that, obviously, fully symmetric states
are bisymmetric under any bipartition of the modes: pictorially (see
Fig.~\ref{figbasset}), they are thus a special type of basset hound
state resembling a {\em Cerberus} state, in which any one of the
three heads  can be cut and can be reversibly regrown. Only in this
case can a promiscuous sharing of entanglement arise. It is worth stressing
that fully symmetric Gaussian states include nearly all the states
of tripartite CV systems currently produced in the laboratory by
quantum optical means \cite{3mexp,pfister}; we will discuss their
usefulness as resources for quantum communication protocols with
continuous variables in Sec. \ref{secpoppy}.

Let us also mention that the argument connecting promiscuous
entanglement to complete permutation invariance does not hold
anymore in the case of Gaussian states with four and more modes,
where relaxing the symmetry constraints allows for an {\em
enhancement} of the distributed entanglement promiscuity to an
unlimited extent \cite{unlim} (thanks to the greater freedom
available in a $8$-dimensional phase space).

\section{Optical production of three-mode Gaussian states}\label{engi}

In this Section, we present a novel (and more economical, in a
well-defined sense, with respect to previous proposals) recipe to
generate pure three-mode Gaussian states with any, arbitrary,
entanglement structure. Moreover, we provide a systematic analysis
of state engineering of the classes of three-mode Gaussian states --
characterized by peculiar structural and/or entanglement properties
-- introduced in the previous section and in Ref.~\cite{3mpra}.  For
every family of Gaussian states, we shall outline schemes for their
production with current optical technology \cite{fabio}.

\subsection{The ``allotment'' box for the production of arbitrary three-mode pure states}\label{secallot}
The investigation of the structural properties and the computation
of the tripartite entanglement, quantified by the residual Gaussian
contangle of \eq{gtaures} (arravogliament), in general {\em pure}
three-mode Gaussian states has been presented in full detail in Ref.~\cite{3mpra}. Here we investigate how to engineer these states with
optical means, allowing for any possible entanglement structure.

A viable scheme
to produce all pure three-mode Gaussian states, as inspired by
Euler decomposition \cite{arvind}, would combine three independent
squeezed modes (with in principle all different squeezing factors)
into any conceivable combination of orthogonal (energy preserving)
symplectic operations (essentially, beam-splitters and
phase-shifters). This procedure, that is obviously legitimate and
will surely be able to generate any pure state, is however not, in
general, the most economical one in terms of physical resources.
Moreover, this procedure is not particularly insightful because
the degrees of bipartite and tripartite entanglement of the resulting
output states is not, in general, easily related to the performed operations.

In this section, we want instead to give a precise recipe providing
the exact operations to achieve a three-mode pure Gaussian state
with any given triplet $\{a_1,\,a_2,\,a_3\}$ of local mixedness, and
so with any desired `physical' ({\ie}, constrained by Inequality
(\ref{triangleprim})) asymmetry among the three modes and any needed
amount of tripartite entanglement. Clearly, such a recipe is {\em
not} unique.\footnote{An alternative scheme to produce pure
three-mode Gaussian states can be inferred from Ref. \cite{generic},
where the state engineering of pure $N$-mode Gaussian states with no
correlations between position and momentum operators is discussed.
We will discuss it in comparison with the present scheme, later in
the text.} We provide here one possible, novel scheme, which may not
be the cheapest one but possesses a straightforward physical
interpretation: the passive distribution, or {\em allotment} of
two-mode entanglement among three modes.

Explicitly, one starts with modes $1$ and $2$ in a two-mode squeezed
state (which can be obtained in the lab
\cite{francamentemeneinfischio}, either directly in non-degenerate
parametric processes or by mixing two squeezed vacua at a
beam-splitter), and mode $3$ in the vacuum state. In Heisenberg
picture:
\begin{eqnarray}
  &\hat q_1 = \frac{1}{\sqrt{2}}\left(e^{r}\ \hat q_1^0+e^{-r}\ \hat q_2^0\right)\,,&\quad
  \hat p_1 =  \frac{1}{\sqrt{2}}\left(e^{-r}\ \hat p_1^0+e^{r}\ \hat p_2^0\right)\,, \label{mode1bas}  \\
    &\hat q_2 = \frac{1}{\sqrt{2}}\left(e^{r}\ \hat q_1^0-e^{-r}\ \hat q_2^0\right)\,,&\quad
  \hat p_2 =  \frac{1}{\sqrt{2}}\left(e^{-r}\ \hat p_1^0-e^{r}\ \hat p_2^0\right)\,, \label{mode2bas}  \\
  &\hat q_3 = \hat q_3^0\,,&\quad
  \hat p_3 =  \hat p_3^0\,, \label{mode3bas}
\end{eqnarray}
where the suffix ``0'' refers to the vacuum.

\begin{figure}[t!]
\centering{
\includegraphics[width=7cm]{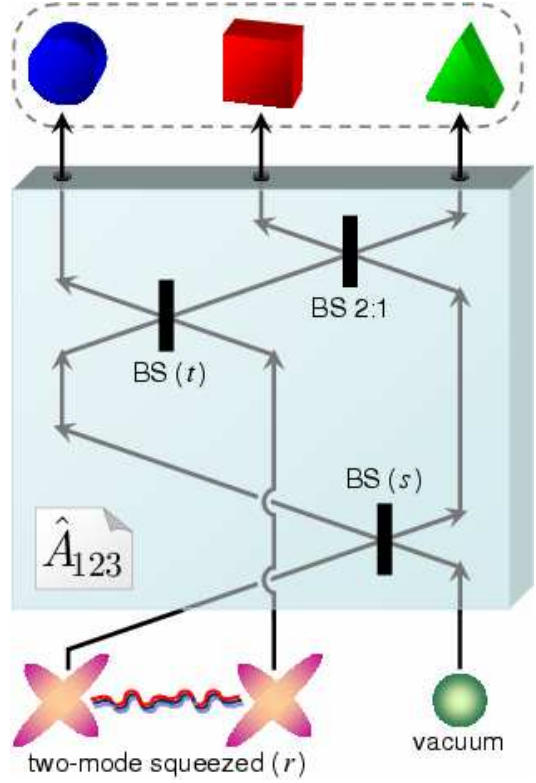}
\caption{Scheme to produce arbitrary pure three-mode Gaussian states
(up to local unitaries). A two-mode squeezed state and a single-mode
vacuum are combined by the ``allotment'' operator $\hat{A}_{123}$,
which is a sequence of three beam-splitters, \eq{allot}. The output
yields an arbitrary pure Gaussian state of modes $1$ (blue circle
\textcolor[rgb]{0.00,0.00,1.00}{\ding{108}}), $2$ (red square
\textcolor[rgb]{1.00,0.00,0.00}{\ding{110}}), and $3$ (green
triangle \textcolor[rgb]{0.00,1.00,0.00}{\ding{115}}), whose CM
depends on the initial squeezing factor $m = \cosh(2r)$ and on two
beam-splitter transmittivities $s$ and $t$.} \label{allocco}}
\end{figure}

The three initial modes are then sent in a sequence of three
beam-splitters, which altogether realize what we will call
``allotment'' operator and denote by $\hat{A}_{123}$ (see Fig.
\ref{allocco}):
\begin{equation}\label{allot}
\hat{A}_{123} \equiv \hat{B}_{23} (\arccos\sqrt{2/3}) \cdot
\hat{B}_{12} (\arccos\sqrt{t}) \cdot \hat{B}_{13}
(\arccos\sqrt{s})\,.
\end{equation}
Here the action of an ideal (phase-free) beam-splitter operation
$\hat{B}_{ij}$ on a pair of modes $i$ and $j$ is defined as
\begin{equation}\label{bsplit}
\hat{B}_{ij}(\theta):\left\{
\begin{array}{l}
\hat a_i \rightarrow \hat a_i \cos\theta + \hat a_j\sin\theta \\
\hat a_j \rightarrow \hat a_i \sin\theta - \hat a_j\cos\theta \\
\end{array} \right.\,,
\end{equation}
with $\hat a_l = (\hat x_l + i \hat p_l)/2$ being the annihilation
operator of mode $k$, and $\theta$ the angle in phase space
($\theta=\pi/4$ corresponds to a 50:50 beam-splitter).

It is convenient in this instance to deal with the phase-space
representations of the states ({\ie}their CM) and of the operators
({\ie}the associated symplectic transformations). The three-mode
input state is described by a CM $\sig^p_{in}$ of the form
\eq{subma} for $n=3$, with
\begin{eqnarray}
  &\sig_1=\sig_2=m\ \id_2\,,\quad&\sig_3=\id_2\,, \\
  &\eps_{12} = {\rm diag}
  \left\{\sqrt{m^2-1},\,-\sqrt{m^2-1}\right\}\,,\quad&\eps_{13}=\eps_{23}= {\bf
  0}\,,
\end{eqnarray}
and $m\equiv \cosh(2r)$. A beam-splitter with transmittivity $\tau$
corresponds to a rotation of $\theta = \arccos\sqrt{\tau}$ in phase
space, see \eq{bsplit}. In a three-mode system, the symplectic
transformation corresponding to $\hat B_{ij}(\theta)$ is a direct
sum of the matrix $B_{ij}(\tau)$,
\begin{equation}\label{bbs}
B_{ij}(\tau)=\left(
\begin{array}{cccc}
 \sqrt{\tau } & 0 & \sqrt{1 - \tau } & 0 \\
 0 & \sqrt{\tau } & 0 & \sqrt{1 - \tau } \\
 \sqrt{1 - \tau } & 0 & - \sqrt{\tau } & 0 \\
 0 & \sqrt{1 - \tau } & 0 & - \sqrt{\tau }
\end{array}
\right)\,,
\end{equation}
acting on modes $i$ and $j$, and of the identity $\id_2$ acting on
the remaining mode $k$.

The output state after the allotment will be denoted by a CM
$\sig_{out}^p$ given by
\begin{equation}\label{aftall}
\sig_{out}^p = A_{123} \sig_{in}^p A_{123}^{\sf T}\,,
\end{equation}
where $A_{123}$ is the phase-space representation of the allotment
operator \eq{allot}, obtained from the matrix product of the three
beam-splitter transformations. The output state is clearly pure
because the allotment is a unitary operator (symplectic in phase
space). The elements of the CM $\sig_{out}^p$,  not reported here
for brevity, are functions of the three parameters
\begin{equation}\label{mst}
m \in [1,\,\infty),\,\quad s \in [0,\,1],\,\quad t \in [0,\,1]\,.
\end{equation}
In fact, by letting these three parameters vary in their
respective domain, the presented procedure allows for the creation
of three-mode pure Gaussian states with any
possible triplet of local mixednesses $\{a_1,\,a_2,\,a_3\}$ ranging
in the physical region defined by the triangle inequality
(\ref{triangleprim}), and thus encompassing all the possible entanglement
structures, under any partition of the system.

This can be shown as follows. Once identified $\sig_{out}^p$ with the
block form of \eq{subma} (for $n=3$), one can solve analytically the
equation $\det \sig_1 = a_1^2$ to find
\begin{equation}\label{mgen}
m(a_1,s,t)=\begin{array}{c} \frac{t \left(t (s - 1)^2 + s - 1\right)
+ \sqrt{a_1^2 (s t + t - 1)^2  +
            4 s (t - 1) t (2 t - 1) (2 s t - 1)}}{(s t + t -
            1)^2}\,.
            \end{array}
\end{equation}
Then, substituting \eq{mgen} in $\sig_{out}^p$ yields a
reparametrization of the output state in terms of $a_1$ (which is
given), $s$ and $t$. Now solve (numerically) the system of nonlinear
equations $\{\det\sig_2 = a_2^2,\,\det\sig_3 = a_3^2\}$ in the
variables $s$ and $t$. Finally, substitute back the obtained values
of the two transmittivities in \eq{mgen}, to have the desired
triplet $\{m,\,s,\,t\}$ as functions of the local mixednesses
$\{a_1,\,a_2,\,a_3\}$ characterizing the target state.

\begin{figure}[t!]
\centering{
\includegraphics[width=8cm]{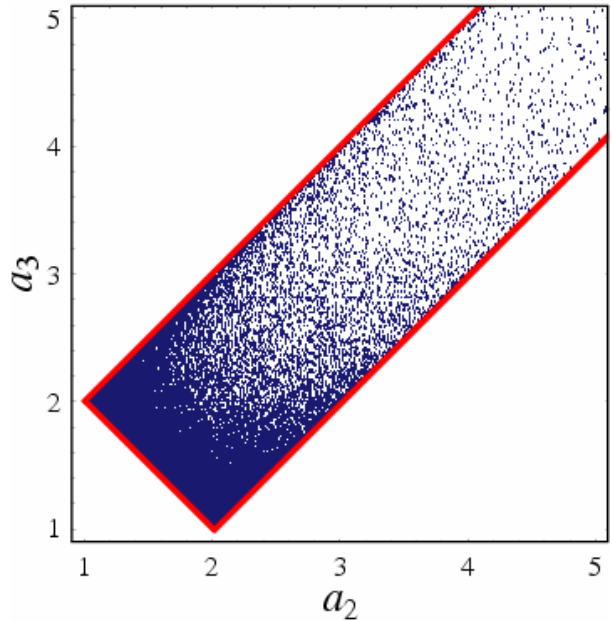}
\caption{Plot of 100000 randomly generated pure three-mode Gaussian
states, described by their single-mode mixednesses $a_2$ and $a_3$,
at fixed $a_1=2$. The states are produced by simulated applications
of the allotment operator with random beam-splitter transmittivities
$s$ and $t$, and span the whole physical range of parameters allowed
by \ineq{triangleprim}. A comparison of this plot with Fig. 1 of
Ref. \cite{3mpra} may be instructive. See text for further details.}
\label{pompilio}}
\end{figure}

An arbitrary pure three-mode Gaussian state, with a CM locally
equivalent to the standard form of \eq{subma} with all diagonal $2
\times 2$ subblocks, can thus be produced with the current
experimental technology by linear quantum optics, employing the
allotment box with exactly tuned amounts of input two-mode squeezing
and beam-splitter properties, without any free parameter left.
The outcome of this procedure is shown in Fig. \ref{pompilio},
where at a given local mixedness of mode $1$ ($a_1=2$), several runs
of the allotment operator have been simulated adopting beam-splitters
of random transmittivities $s$ and $t$. Starting from a two-mode
squeezed input with $m$ given by \eq{mgen}, tensor a vacuum, the
resulting output states are plotted in the space of $a_2$
and $a_3$. By comparing Fig. \ref{pompilio} with Fig. 1 of Ref.
\cite{3mpra}, one sees clearly that the randomly generated states
distribute towards a complete fill of the physical region emerging
from the triangle inequality (\ref{triangleprim}), thus confirming
the generality of our scheme.

A remark is in order here. The optical scheme presented in Ref.~\cite{generic}
to produce pure $N$-mode Gaussian states with
``generic entanglement'' (corresponding to standard form CMs with
null position-momentum covariances), also allows in the special case
of $N=3$ for the creation of {\em all} pure three-mode Gaussian
states with CM in standard form (see Sec. \ref{struct} and Ref.
\cite{3mpra}).
Therefore, such a state engineering recipe represents an alternative
to the allotment box of Fig.~\ref{allocco}. However, there is a
crucial difference between the two schemes: the allotment box
requires {\it in toto} (considering both the preparation and the
further manipulation of the input states out of three vacuum beams)
a single ``active'' (\ie non energy-preserving), squeezing
operation, whereas the scheme of Ref.~\cite{generic} needs two
squeezing operations to be accomplished. In this specific sense, the
allotment strategy is ``more economical'' over the scheme of
Ref.~\cite{generic} (in view of the superior expediency with which
passive operations -- beam-splitters, in this instance -- can be
realized in practice, being considerably more efficient and reliable
than squeezings). More in detail, in both schemes, the input in
modes 1 and 2 is a two-mode squeezed state, whose squeezing
parameter accounts for one of the three degrees of freedom of pure
three-mode Gaussian states in standard form. Still, while in the
present scheme mode 3 starts off in the vacuum state, the scheme of
Ref.~\cite{generic} requires an initial single-mode squeezed state
in mode 3: a single beam-splitter (between modes 2 and 3) is then
enough to achieve a completely general entanglement structure. On
the other hand, the allotment box presented here is realized by a
{\em passive} redistribution of entanglement only, as the third mode
is not squeezed, but the three modes need to interfere with each
other via three beam-splitters (one of which has fixed
transmittivity) and this again yields a completely general
entanglement freedom. Summing up, the allotment scheme presented
here needs four operations, three of which are passive, while the
scheme of Ref.~\cite{generic} needs two active operations and one
passive operation to achieve full generality. Depending on the
specific experimental facilities, one may thus choose either scheme
when aiming to produce pure three-mode Gaussian states. Clearly, if
a very high amount of final entanglement is required, the strategy
adopting {\em two} squeezing operations in lieu of one might be more
suitable  (as the squeezing achievable in a single parametric
process is limited); however, such larger entanglement will come at
the price of a considerably higher level of noise (as parametric
processes are generally less stable than passive linear optics).

\subsection{Concise guide to tripartite state engineering and simplified schemes}\label{sectstates}

Let us now turn to the production of the classes of three-mode
states introduced in Sec.~\ref{promo} and in Ref.~\cite{3mpra}. For
mixed instances of such tripartite states (not subsumed by the
allotment operator), efficient state engineering schemes will be
outlined. Also, in special instances of pure states, depending in
general on less than three parameters, cheaper recipes than the
general one in terms of the allotment box will be presented.

\subsubsection{Pure and noisy GHZ/{\it W} states}\label{SecEngiGHZW}

\begin{figure}[t!]
\centering{
\includegraphics[width=8cm]{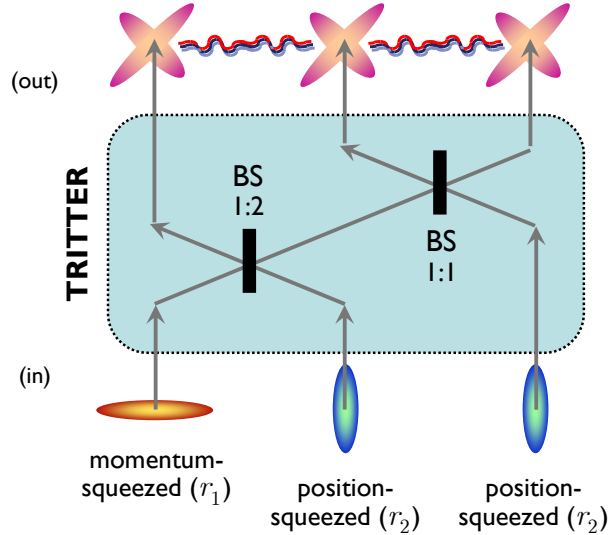}
\caption{Scheme to produce CV GHZ/$W$ states, as proposed in
Ref.~\cite{network} and implemented in Ref.~\cite{3mexp}. Three
independently squeezed beams, one in momentum and two in position,
are combined through a double beam-splitter (tritter). The output
yields a pure, symmetric, fully inseparable three-mode Gaussian
state, also known as CV GHZ/$W$ state.} \label{fighzw}}
\end{figure}

Several schemes have been proposed to produce pure GHZ/$W$ states
with finite squeezing \cite{contangle,3mpra}, \ie fully symmetric
pure three-mode Gaussian states (with promiscuous entanglement
sharing). In particular, as discussed by Van Loock and Braunstein
\cite{network}, these states can be produced by mixing three
squeezed beams (one in momentum and the other two in position) in a
double beam-splitter, or {\em tritter} \cite{branature}.

One starts with mode $1$ squeezed in momentum, and modes $2$ and $3$
squeezed in position. In Heisenberg picture:
\begin{eqnarray}
  &\hat q_1 = e^{r_1}\ \hat q_1^0\,,&\quad
  \hat p_1 =  e^{-r_1}\ \hat p_1^0\,, \label{momsq3}  \\
   &\hat q_{2,3} =  e^{-r_2}\ \hat q_{2,3}^0\,,&\quad
  \hat p_{2,3} =  e^{r_2}\ \hat p_{2,3}^0\,, \label{possq3}
\end{eqnarray}
where the suffix ``0'' refers to the vacuum. Then one combines the
three modes in a tritter
\begin{equation}\label{tritter}
\hat{B}_{123} \equiv \hat{B}_{23}(\pi/4)\cdot\hat{B}_{12}
(\arccos\sqrt{1/3})\,,
\end{equation}
where the action of an ideal (phase-free) beam-splitter operation
$\hat{B}_{ij}$ on a pair of modes $i$ and $j$ is defined by
\eq{bsplit}.

 The output of the tritter yields a CM of the form
\eq{fscm} with
\begin{eqnarray}
\alp = {\rm diag} \left\{
 \frac{1}{3} \left(e^{2 r_1} + 2 e^{-2 r_2}\right)\,,\quad
 \frac{1}{3} \left(e^{-2 r_1} + 2 e^{2 r_2}\right)\right\}\,, \label{sighzwa}\\
\eps = {\rm diag} \left\{
 \frac{1}{3} \left(e^{2 r_1} - e^{-2 r_2}\right)\,,\quad
\frac{1}{3} \left(e^{-2 r_1} - e^{2 r_2}\right)\right\}\,.
\label{sighzwe}
\end{eqnarray}
This resulting pure and fully symmetric three-mode Gaussian state,
obtained in general with differently squeezed inputs $r_1 \neq r_2$,
is locally equivalent to the state prepared with all initial
squeezings equal to the average $\bar r = (r_1 + r_2)/2$
\cite{telepoppate}.
 The CM described by
Eqs.~{\rm(\ref{sighzwa},\ref{sighzwe})} represents a CV GHZ/$W$
state. It can be in fact transformed, by local symplectic
operations, into the standard form CM of Ref.~\cite{3mpra}, which
obeys \eq{fscm} with $a \equiv \sqrt{\det\gr\alpha}$ given by
\begin{equation}\label{aghzw}
a=\frac{1}{3} \sqrt{4 \cosh \left[2 \left(r_1 + r_2\right)\right] +
5}\,.
\end{equation}

Noisy GHZ/$W$ states, whose entanglement has been characterized in
Sec.~\ref{secnoisyghzw}, can be obviously obtained by an analogous
procedure starting from (Gaussian) thermal states instead of vacua
(with average photon number $\bar n =[n-1]/2\ge0$) and combining
them through a tritter \eq{tritter}. The initial single, separable,
modes are thus described by the following operators in Heisenberg
picture (we will now assume the same squeezing parameter $\bar r
\equiv r$ for the three beams, as it allows for any case of noisy
GHZ/$W$ states up to local unitaries)
\begin{eqnarray}
  &\hat q_1 = \sqrt{n} e^{r}\ \hat q_1^0\,,&\quad
  \hat p_1 =  \sqrt{n} e^{-r}\ \hat p_1^0\,, \label{momthsq}  \\
   &\hat q_{2,3} =  \sqrt{n} e^{-r}\ \hat q_{2,3}^0\,,&\quad
  \hat p_{2,3} =  \sqrt{n} e^{r}\ \hat p_{2,3}^0\,. \label{posthsq}
\end{eqnarray}
Defining $s \equiv e^{2r}$, at the output of the tritter one obtains
a CM of the form \eq{fscm}, with
\begin{eqnarray}
\alp = {\rm diag} \left\{
 \frac{n(s^2+2)}{3s}\,,\quad
 \frac{n(2s^2+1)}{3s}\right\}\,, \label{sitha}\\
\eps = {\rm diag} \left\{
 \frac{n(s^2-1)}{3s}\,,\quad
-\frac{n(s^2-1)}{3s}\right\}\,. \label{sithe}
\end{eqnarray}
This resulting CM is locally equivalent to the standard form of of
Ref.~\cite{3mpra}, which obeys \eq{fscm} with $a \equiv
\sqrt{\det\gr\alpha}$ given by
\begin{equation}\label{athermal}
a = \frac{n \sqrt{2 s^4 + 5 s^2 + 2}}{3 s}\,.
\end{equation}
Clearly, setting $n=1$ corresponds to the case of pure
GHZ/$W$ states.

The preparation scheme of CV GHZ/$W$ states is depicted in
Fig.~\ref{fighzw}. It has has been experimentally implemented
\cite{3mexp}, and the full inseparability of the produced states has
been verified through the violation of the separability inequalities
derived in Ref.~\cite{vanlokfuru}. Very recently, the production of
strongly entangled GHZ/$W$ states has also been demonstrated by
using a novel optical parametric oscillator, based on concurrent
$\chi^{(2)}$ nonlinearities \cite{pfister}.

\subsubsection{{\it T} states}\label{SecEngiT}

The $T$ states have been introduced in Refs.~\cite{3mpra,contangle}
to show that, in symmetric three-mode Gaussian states, imposing the
absence of reduced bipartite entanglement between any two modes
results in a frustration of the genuine tripartite entanglement. It
may be useful to know how to produce this novel class of mixed,
fully symmetric Gaussian states in the lab.

The simplest way to engineer $T$ states is to reutilize the scheme
of Fig.~\ref{fighzw}, {\ie}basically the tritter of \eq{tritter},
but with different inputs. Namely,  one has mode $1$ squeezed again
in momentum (with squeezing parameter $r$), but this time modes $2$
and $3$ are in a thermal state (with average photon number $\bar n =
[n(r)-1]/2$, depending on $r$). In Heisenberg picture:
\begin{eqnarray}
  &\hat q_1 = e^{r}\ \hat q_1^0\,,&\quad
  \hat p_1 =  e^{-r}\ \hat p_1^0\,, \label{momsqt}  \\
   &\hat q_{2,3} =  \sqrt{n(r)}\ \hat q_{2,3}^0\,,&\quad
  \hat p_{2,3} =  \sqrt{n(r)}\ \hat p_{2,3}^0\,, \label{thermt}
\end{eqnarray}
with $n(r) = \sqrt{3 + e^{-4 r}} - e^{-2 r}$. Sending these three
modes in a tritter \eq{tritter} one recovers, at the output, a $T$
state whose CM is locally equivalent to the standard form of
Ref.~\cite{3mpra}, with
\begin{equation}\label{atstat}
a = \frac{1}{3} \sqrt{2 e^{-2 r} \sqrt{3 + e^{-4 r}} \left(-3 + e^{4
r}\right) + 6 e^{-4 r} + 11}\,.
\end{equation}

\subsubsection{Basset hound states}\label{secbasengi}

A scheme for producing the basset hound states of Sec.~\ref{secbas},
and in general the whole family of pure bisymmetric Gaussian states
known as ``multiuser quantum channels'' (due to their usefulness for
telecloning, as we will show in Sec.~\ref{sectlc}), is provided in
Ref.~\cite{telecloning}. In the case of three-mode pure basset hound
states of the form given by Eqs.~{\rm(\ref{bassigl},\ref{basseps})},
one can use a simplified version of the allotment introduced in
Sec.~\ref{secallot} for arbitrary pure states. One  starts with a
two-mode squeezed state (with squeezing parameter $r$) of modes $1$
and $2$, and mode $3$ in the single-mode vacuum,  like in
Eqs.~{\rm(\ref{mode1bas}--\ref{mode3bas})}. Then, one combines one
half (mode 2) of the two-mode squeezed state  with the vacuum mode 3
via a 50:50 beam-splitter, described in phase space by $B_{23}(1/2)$
of \eq{bbs}. The resulting three-mode state is exactly a basset
hound state described by Eqs.~{\rm(\ref{bassigl},\ref{basseps})},
once one identifies
$a \equiv \cosh(2r)$.
In a realistic setting, dealing with noisy input modes, mixed
bisymmetric states can be obtained as well by the same procedure.

\section{Application: Multiparty quantum teleportation with continuous variables}\label{intrappli}

We now address more closely the usefulness of three-mode Gaussian
states for the efficient implementation of quantum information and
communication protocols. In particular, we intend to provide the
theoretical entanglement characterization of three-mode Gaussian
states with a significant operative background and to epitomize
their capability for quantum communication tasks. To this aim, we
shall focus on the transmission of quantum states within a network
of three parties which share entangled Gaussian resources.

For two parties, the
 process of {\em quantum teleportation} using  entanglement and with the aid of
 classical communication was originally proposed for qubit systems \cite{telep}, and
experimentally  implemented with polarization-entangled photons
\cite{telezei,teledem}. The CV counterpart of discrete-variable
teleportation, using quadrature entanglement, is in principle
imperfect due to the impossibility of achieving infinite squeezing.
Nevertheless, by considering the finite EPR correlations between the
quadratures of a two-mode squeezed Gaussian state, a realistic
scheme for CV teleportation (see Ref. \cite{pirandolareview} for a
recent review) was proposed \cite{vaidman,bratele} and
experimentally implemented \cite{furuscience} to teleport coherent
states with a measured fidelity ${\cal F} = 0.70 \pm 0.02$
\cite{furunew}. Without using entanglement, by purely classical
communication, an average fidelity of
\begin{equation}\label{fcl}
{\cal F}_{cl}=\frac{1}{2}
\end{equation}
is the maximal achievable with an alphabet of uniformly distributed coherent input
states \cite{bfkjmo,hammerer}.
Let us recall that the fidelity ${\cal F}$, which is the figure of merit quantifying the
success of a teleportation experiment, is defined with respect to a pure state $\ket{\psi^{in}}$ as
\begin{equation}\label{fid}
{\cal F} \equiv \bra{\psi^{in}} \varrho^{out}\ket{\psi^{in}}\, .
\end{equation}
Here ``in'' and ``out'' denote the input and the output states (the
latter being generally mixed) of a teleportation process,
respectively. ${\cal F}$ reaches unity only for a perfect state
transfer, $\varrho^{out} = \ket{\psi^{in}}\!\bra{\psi^{in}}$. To
accomplish teleportation with high fidelity, the sender (Alice) and
the receiver (Bob) must share an entangled state (resource). The
{\em sufficient} fidelity criterion \cite{bfkjmo} states that, if
teleportation is performed with ${\cal F} > {\cal F}_{cl}$, then the
two parties exploited an entangled state. The converse is generally
false: that is, quite surprisingly, some entangled resources may
yield lower-than-classical fidelities \cite{telepoppate,network}.
This point will be discussed thoroughly in the following.

To generalize the process of CV teleportation from two to three (and
more) users, one can consider two basic possible scenarios.
On the one hand, a
network may be created where each user is able to
teleport states with better-than-classical efficiency (being the
same for all sender/receiver pairs) to any chosen receiver {\em with the
assistance of the other parties}. On the other hand, one of the
parties may act as the fixed sender, and distribute approximate
copies (with in principle different cloning fidelities) to all the
others acting as remote receivers. These two protocols, respectively
referred to as ``teleportation network'' \cite{network} and
``telecloning'' \cite{telecloning}, will be described in the two
following sections, and the connections between their successful
implementation with three-mode Gaussian resources and the amounts of
shared bipartite and tripartite entanglement will be elucidated.
We just mention that several interesting variants to these basic schemes do exist
(see, {\em e.g.} the `cooperative telecloning' of Ref.~\cite{pirgame}, where
two receivers -- instead of two senders -- are cooperating).

\section{Teleportation networks with fully symmetric resources} \label{secpoppy}

The original CV teleportation protocol \cite{bratele} has been
generalized to a multi-user teleportation network requiring
multiparty entangled Gaussian states in Ref.~\cite{network}. This
network has been recently experimentally demonstrated by exploiting
three-mode squeezed  states (namely, noisy CV GHZ/W states,
extensively addressed in section \ref{secnoisyghzw}, were employed),
yielding a maximal fidelity ${\cal F} = 0.64 \pm 0.02$
\cite{naturusawa}.

In Ref. \cite{telepoppate} the problem was raised of determining the
{\em optimal} multi-user teleportation fidelity (in the general
$N$-mode setting), and to extract from it a quantitative information
on the multipartite entanglement in the shared resource. The
optimization consists in a maximization of the fidelity over all
local single-mode operations (`pre-processing' the initial
resource), at fixed amounts of noise and entanglement in the shared
resource. This is motivated by the simple observation that states
equivalent up to local single-mode unitary operations (which
possess, by definition, the same amount of bipartite and
multipartite entanglement with respect to any partition) behave in
general differently when employed in a fixed (not locally optimized)
quantum information protocol.\footnote{Clearly, this fact is, {\em
per se}, not that surprising. Also, it may be noted that one could,
having a complete knowledge about the resource, always include the
local optimizing pre-processing as the first step of the considered
``protocol''.} In the previous section, dedicated to state
engineering, we provided schemes for the generation of states with
CMs {\em locally equivalent} to the corresponding standard forms.
The teleportation efficiency, instead, depends separately on the
different single-mode properties (in particular, on the squeezings
degrees of the input states before combining them via optical
networks like the tritter or the allotment).

 For fully
symmetric (pure or mixed) shared resource, it has been shown in
Ref.~\cite{telepoppate} that
the optimal fidelity ${\cal F}^{opt}_N$ obtained in such way is {\em
equivalent} to the presence of genuine multipartite entanglement in
the shared resource. This results yield quite naturally a direct
operative way to quantify multipartite entanglement in $N$-mode
(mixed) symmetric Gaussian states, in terms of the so-called {\em
Entanglement of Teleportation}, defined as the normalized optimal
fidelity
\begin{equation}\label{et}
E_T^{(N)} \equiv \max\left\{0,\frac{{\cal F}_N^{opt}-{\cal
F}_{cl}}{1-{\cal F}_{cl}}\right\}\,,
\end{equation} with ${\cal F}_{cl}\equiv 1/2$ being the classical threshold
For any $N$, the entanglement of teleportation ranges from 0
(separable resource states) to 1 (CV generalized GHZ resource state,
simultaneous eigenstate of total momentum and all relative positions
of the $N$-mode radiation field).
As we will discuss in detail later on, the equivalence between optimal fidelity of teleportation
and entanglement breaks down in the asymmetric instance, even for
two-mode states.

We shall now remind a particular relationship (first found in Ref.~\cite{telepoppate})
between
the entanglement of teleportation
and the residual Gaussian contangle introduced in
Sec.~\ref{sepa2m} and further discuss its operational consequences in terms of teleportation networks,
eventually leading to a proposal to experimentally test the promiscuous sharing of correlations with three-mode
Gaussian states.

\subsection{On the operational interpretation of tripartite Gaussian
entanglement and on how to experimentally investigate its sharing
structure}

\subsubsection{Entanglement of teleportation and residual contangle}

Let us focus, for the following discussion, on the case $N=3$,
{\ie}on three-mode states shared as resources for a three-party
teleportation network. This protocol is a basic, natural candidate
to operationally investigate the sharing structure of CV
entanglement in three-mode symmetric Gaussian states.

A first theoretical question that arises is to compare the
tripartite entanglement of teleportation \eq{et}, which is endowed
with a strong operational motivation \cite{telepoppate}, and the
tripartite residual (Gaussian) contangle \eq{gtaures}, which is
built on solid mathematical foundations. Remarkably, in the case of
{\em pure} and {\em symmetric} three-mode resources ({\ie}for CV
GHZ/$W$ states) the two measures are completely equivalent
\cite{telepoppate}, being monotonically increasing functions of each
other. Namely, from \eq{gresghzw},
\begin{equation}\label{etetau} \hspace*{-2cm}
G_\tau^{res}(\sig_{s}^{_{{\rm GHZ}/W}}) = \ln^2{\frac{2\sqrt2
E_T-(E_T+1)\sqrt{E_T^2+1}}{(E_T-1)\sqrt{E_T(E_T+4)+1}}}-\frac12
\ln^2{\frac{E_T^2+1}{E_T(E_T+4)+1}}\,,
\end{equation}
where $E_T \equiv E_T^{(3)}$ in \eq{et}. Let us moreover recall that
$G_\tau^{res}$ coincides with the true residual contangle (globally
minimized in principle over all, including non-Gaussian,
decompositions), \eq{etaumin}, in these states
\cite{contangle,3mpra}.

Therefore, in the specific (but relevant) instance of symmetric pure states,
the residual (Gaussian) contangle is enriched of an interesting
meaning as a {\em resource} enabling a better-than-classical
three-party teleportation experiment, while no operational
interpretations are presently known for the three-way residual
tangle quantifying tripartite entanglement sharing in qubit systems
\cite{CKW,wstates}. We remark that in the tripartite instance, the
optimal fidelity ${\cal F}_{3}^{opt}$ -- determining the entanglement of teleportation --
achieves indeed its {\em
global} maximum over all possible Gaussian POVMs performed on the
shared resource, as can be confirmed with the methods of
Ref.~\cite{pirandolassisted}.

\subsubsection{The role of promiscuity in symmetric three-mode
resources}\label{secpromis}

 The relationship between optimal teleportation fidelity and
residual (Gaussian) contangle, embodied by \eq{etetau},  entails
that there is a `unique' kind of three-party CV entanglement in pure
{\em symmetric} three-mode Gaussian states (alias CV
finite-squeezing GHZ/$W$ states), which merges at least three
(usually inequivalent) properties: those of being maximally
genuinely tripartite entangled, maximally bipartite entangled in any
two-mode reduction, and `maximally efficient' (in the sense of the
optimal fidelity) for three-mode teleportation networks. Recall that
the first two properties, taken together, label such entanglement as
{\em promiscuous}, as discussed in Sec.~\ref{secpromis}. These
features add up to the property of tripartite GHZ/$W$ Gaussian
states of being maximally robust against decoherence effects among
all three-mode Gaussian states, as shown in Ref.~\cite{3mpra} and
operatively demonstrated later in Sec.~\ref{qnoise}.

All this theoretical evidence strongly promotes GHZ/$W$ states,
experimentally realizable with current optical technology
\cite{3mexp,pfister} (see Sec.~\ref{SecEngiGHZW}), as paradigmatic
candidates for the encoding and transmission of CV quantum
information and in general for reliable CV quantum communication.
 Let us mention that, in particular, these tripartite entangled symmetric Gaussian states
have been successfully employed to demonstrate quantum secret
sharing \cite{secret}, controlled dense coding \cite{dense}, and the
above discussed teleportation network \cite{naturusawa}. Quite recently, a
theoretical solution for CV Byzantine agreement has been reported
\cite{sanpera}, based on the use of sufficiently entangled states
from the family of CV GHZ/$W$ states.

Building on our entanglement analysis, we can precisely enumerate
the peculiarities of those states which make them so appealing for
practical implementations.  Exploiting a strongly entangled
three-mode CV GHZ/$W$ state as a quantum channel affords one with a
number of simultaneous advantages:
\begin{enumerate}\label{checklist}
  \item[{(i)}] the ``guaranteed success'' ({\ie}with better-than-classical
figures of merit) of any known tripartite CV quantum information
protocol;
  \item[{(ii)}] the ``guaranteed success'' of any standard two-user CV
protocol, because a highly entangled two-mode channel is readily
available after a unitary (reversible) localization of entanglement
has been performed through a single beam-splitter (see
Fig.~\ref{figbasset});
  \item[{(iii)}] the ``guaranteed success'' (though with nonmaximal
efficiency) of any two-party quantum protocol through each two-mode
channel obtained discarding one of the three modes.
\end{enumerate}
Point (iii) ensures that, even when one mode is lost, the remaining
(mixed) two-mode resource can be still implemented for a two-party
protocol with better-than-classical success. It is realized with
 nonmaximal efficiency because, as we have seen from
\eq{gredmaxghzw}, the reduced entanglement in any two-mode partition
remains finite even with infinite squeezing (the reason why
promiscuity of tripartite Gaussian entanglement is only ``partial'',
compared to the four-partite case of Ref.~\cite{unlim}).

We can now readily provide an explicit proposal to implement the
above checklist in terms of CV teleportation networks.

\subsubsection{Testing the promiscuous sharing of tripartite
entanglement}\label{sectest} 
\begin{figure}[t!]
\centering{\includegraphics[width=13cm]{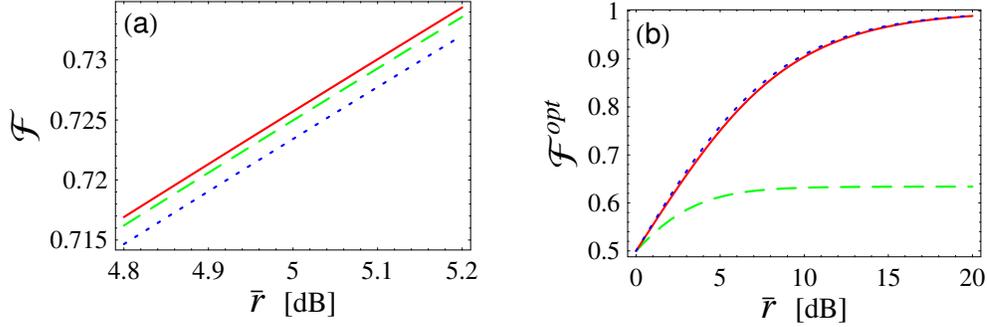}%
\caption{\label{figapoppa} \textsf{(a)} Plot of the fidelities for
teleporting an arbitrary coherent state from any sender to any
receiver chosen from $N=3$ parties, sharing a GHZ/$W$ state. In a
small window of average squeezing, we compare the optimal fidelity
\cite{telepoppate} (solid line), the fidelity obtained for the
unbiased states discussed in Ref. \cite{bowen} (dashed line), and
the fidelity for states produced with all equal squeezers
\cite{network} (dotted line). The three curves are very close to
each other, but the optimal preparation yields always the highest
fidelity, as first proven in \cite{telepoppate}. \textsf{(b)}
Expected success for an experimental test of the promiscuous sharing
of CV entanglement in GHZ/$W$ states. Referring to the check-list in
Sec. \ref{secpromis}: the solid curve realizes point (i), being the
optimal fidelity ${\cal F}_3^{opt}$ of a three-party teleportation
network; the dotted curve realizes point (ii), being the optimal
fidelity ${\cal F}^{opt}_{2:uni}$ of two-party teleportation
exploiting the two-mode pure resource obtained from a unitary
localization applied on two of the modes; the dashed curve realizes
point (iii), being the optimal fidelity ${\cal F}^{opt}_{2:red}$ of
two-party teleportation exploiting the two-mode mixed resource
obtained discarding a mode. All of them lie above the classical
threshold ${\cal F}^{cl}=0.5$, providing a direct evidence of the
promiscuity of entanglement sharing in the employed resources.}}
\end{figure}

The results just elucidated pave the
way towards an experimental test for the promiscuous sharing of
tripartite CV entanglement in symmetric Gaussian states, as
anticipated in the outlook of Ref.~\cite{telepoppate}.To unveil this
peculiar feature, one should prepare a pure CV GHZ/$W$ state
 according to Fig.~\ref{fighzw}, in the optimal form  given by
Ref.~\cite{telepoppate}. It is worth remarking that, in the case of
three modes, non-optimal forms like that produced with $r_1=r_2$ in
Eqs.~(\ref{momsq3}, \ref{possq3}) \cite{3mexp,naturusawa} yield
fidelities really close to the maximal one [see Fig.
\ref{figapoppa}(a)], and are thus practically as good as the optimal
states (if not even better, taking into account that the states with
$r_1=r_2$ are generally easier to produce in practice).

To detect the presence of tripartite entanglement, one should be
able to implement the network in at least two different combinations
\cite{naturusawa}, so that the teleportation would be accomplished,
for instance, from mode $1$ to mode $2$ with the assistance of mode
$3$, and from mode $2$ to mode $3$ with the assistance of mode $1$.
To be complete (even if it is not strictly needed
\cite{vanlokfuru}), one could also realize the transfer from mode
$3$ to mode $1$ with the assistance of mode $2$. Taking into account
a realistic asymmetry among the modes, the minimum experimental
fidelity ${\cal F}_3^{opt}$ over the three possible situations would
provide a direct quantitative measure of tripartite entanglement,
through Eqs.~(\ref{et}--\ref{etetau}).

To demonstrate the promiscuous sharing, one would then need to
discard each one of the modes at a time, and perform standard
two-user teleportation between the remaining pair of parties. The
optimal fidelity for this two-user teleportation (which is achieved
exactly for $r_1=r_2 \equiv \bar r$) is
\begin{equation}\label{f2red}
{\cal F}^{opt}_{2:red} = \frac{3}{3+\sqrt{3+6e^{-4 \bar r}}}\,.
\end{equation}
Again, one should implement the three possible configurations and
take the minimum fidelity as figure of merit. As anticipated in
\ref{secpromis}, this fidelity cannot reach unity because the
entanglement in the shared mixed resource remains finite, and in
fact ${\cal F}^{opt}_{2:red}$ saturates to $3/(3+\sqrt{3}) \approx
0.634$ in the limit of infinite squeezing.

Finding simultaneously both ${\cal F}_3^{opt}$ and ${\cal
F}^{opt}_{2:red}$ above the classical threshold \eq{fcl}, at fixed
squeezing $\bar r$, would be a clear experimental fingerprint
of the promiscuous sharing of CV entanglement.
Theoretically, this is true for all $\bar r > 0$, as
shown in Fig. \ref{figapoppa}(b). From an experimental point of
view, the tripartite teleportation network has been recently
implemented, and the genuine tripartite shared entanglement
unambiguosly demonstrated by obtaining a nonclassical teleportation
fidelity (up to $0.64 \pm 0.02$) in all the three possible user
configurations \cite{naturusawa}. Nevertheless, a nonclassical
fidelity ${\cal F}_{2:red}$ in the teleportation exploiting any
two-mode reduction was not observed.

This fact can be consistently explained by taking into account
experimental noise. In fact, even if the desired resource states
were pure GHZ/$W$ states, the unavoidable effects of decoherence and
imperfections resulted in the experimental production of {\em mixed}
states, namely of the noisy GHZ/$W$ states discussed in Sec.
\ref{secnoisyghzw}. It is very likely that the noise was too high
compared with the pumped squeezing, so that the actual produced
states were still fully inseparable, but laid outside the region of
promiscuous sharing (see Fig. \ref{figacunt}), having no
entanglement left in the two-mode reductions. However, increasing
the degree of initial squeezing, and/or reducing the noise sources
might be accomplished with the state-of-the-art equipment employed
in the experiments of Ref.~\cite{naturusawa} (see also
\cite{furunew}). The conditions required for a proper test (to be
followed by actual practical applications) of the promiscuous
sharing of CV entanglement in symmetric three-mode Gaussian states,
as detailed in Sec.~\ref{secpromis}, should be thus met shortly. As
a final remark, let us observe that repeating the same experiment
but employing $T$ states, described in Sec. \ref{sectstates}, as
resources, would be another interesting option. In fact, in this
case the expected optimal fidelity is strictly smaller than in the
case of GHZ/$W$ states, confirming the promiscuous structure in
which the reduced bipartite entanglement enhances the value of the
genuine tripartite one.

With the same GHZ/$W$ shared resources (but also with all symmetric
and bisymmetric three-mode Gaussian states, including $T$ states
\cite{3mpra}, noisy GHZ/$W$ states and basset hound states), one may
also test the power of the unitary localization of entanglement
\cite{unitarily} (see Fig.~\ref{figbasset}), as opposed to the
nonunitary localization of entanglement by measurements
\cite{localizprl,localizpra}, needed for the teleportation network.
Suppose that the three parties Alice, Bob and Claire share a GHZ/$W$
state. If Bob and Claire are allowed to cooperate (nonlocally), they
can combine their respective modes at a 50$:$50 beam-splitter. The
result is an entangled state shared by Alice and Bob, while Claire
is left with an uncorrelated state. The optimal fidelity of standard
teleportation from Alice to Bob with the unitarily localized
resource, reads
\begin{equation}\label{f2uni}
{\cal F}^{opt}_{2:uni} = \left[\frac{1}{3} \left(\sqrt{4 \cosh (4
\bar r) + 5} - 2 \sqrt{\cosh (4 \bar r) - 1}\right) +
1\right]^{-1}\,.
\end{equation}
Notice that ${\cal F}_{2:uni}$ is  larger than ${\cal F}_3^{opt}$
[see Fig. \ref{figapoppa}(b)]. This is true for any number $N$ of
modes, and the difference between the two, at fixed squeezing,
increases with $N$, confirming that the unitarily localizable
entanglement of \cite{unitarily} is strictly stronger than the
(nonunitarily) localizable entanglement of Refs.
\cite{localizprl,localizpra}, as discussed in Ref. \cite{adebook}.
This is of course not surprising, as the unitary localization generally requires
a high degree of nonlocal control on the two subset of modes, while
the localizable entangement is defined in terms of LOCC alone.

\subsection{Degradation of teleportation efficiency under quantum noise}\label{qnoise}

In Ref.~\cite{3mpra} we have addressed the decay of three-partite
entanglement (as quantified by the residual Gaussian contangle) of
three-mode states in the presence of losses and thermal noise. We
aim now at relating such an `abstract' analysis to precise
operational statements, by investigating the decay of the optimal
teleportation fidelity of shared three-mode resources subject to
environmental decoherence. This study will also provide further
heuristic justification for the residual contangle as a proper
measure of tripartite entanglement even for mixed (`decohered')
Gaussian states.  We will focus on the decay of the teleportation
efficiency under decoherence affecting the resource states {\em
after} their distribution to the distant parties.

We will assume, realistically, a local decoherence ({\em i.e.}~with
no correlated noises) for the three modes, in thermal baths with
equal average photon number $n$. The evolving states maintain their
Gaussian character under such evolution (for a detailed description
of the master equation governing the system and of its Gaussian
solutions, refer to \cite{3mpra}).

\begin{figure}[t!]
\centering{
\includegraphics[width=10cm]{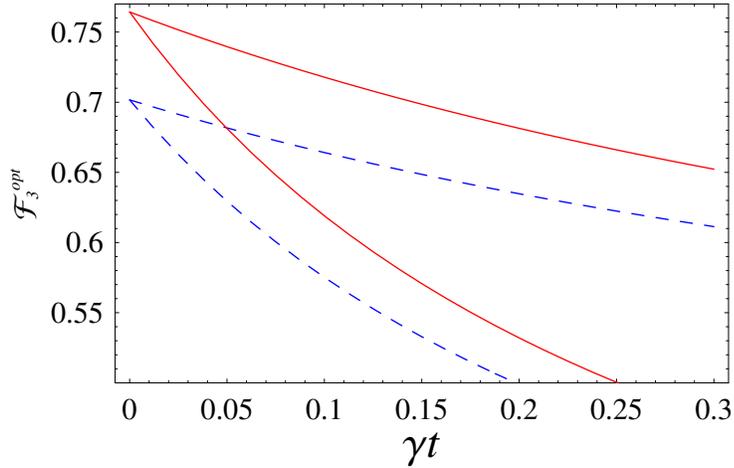}
\caption{Evolution of the optimal fidelity ${\cal F}^{opt}_{3}$ for
GHZ/$W$ states with local mixedness $a=2$ (corresponding to $\bar{r}
\simeq 0.6842$) (solid curves) and $T$ states with local mixedness
$a=2.8014$ (dashed curves). Such states have equal initial residual
contangle but allow for different initial fidelities.
The uppermost curves refer to baths with $n=0$ (`pure
losses'), while the lowermost curves refer to baths with $n=1$. $T$
states affording for the same initial fidelity as the considered
GHZ/$W$ state were also considered, and found to degrade faster than
the GHZ/$W$ state.} \label{teledeco}}
\end{figure}

As initial resources, we have considered both pure GHZ/$W$ states
and mixed $T$ states, as described in Sec.~\ref{sectstates}. The
results, showing the exact evolution of the fidelity ${\cal
F}^{opt}_{3}$ (optimized over local unitaries) of teleportation
networks exploiting such initial states, are shown in
Fig.~\ref{teledeco}. GHZ/$W$ states, already introduced as
``optimal'' resources for teleportation networks, were also found to
allow for protocols most robust under decoherence. Notice how the
qualitative behaviour of the curves of Fig.~\ref{teledeco} follow
that of Fig.~5 of Ref.~\cite{3mpra}, where the evolution of the
residual Gaussian contangle of the same states under the same
conditions is plotted. Also the vanishing of entanglement at finite
times (occurring only in the presence of thermal photons, {\em
i.e.}~for $n>0$) reciprocates the fall of the fidelity below the
classical threshold of $1/2$. The status of the residual Gaussian
contangle as a measure reflecting operational aspects of the states
is thus strengthened in this respect, even in the region of mixed
states. Notice, though, that Fig.~\ref{teledeco} also shows that the
entanglement of teleportation is {\em not} in general quantitatively
equivalent (but for the pure-state case) to the residual Gaussian
contangle, as the initial GHZ/$W$ and $T$ states of
Fig.~\ref{teledeco} have the same initial residual Gaussian
contangle but grant manifestly different fidelities and, further,
the times at which the classical threshold is trespassed do not
exactly coincide with the times at which the residual contangle
vanishes.

This confirms the special role of pure fully symmetric GHZ/$W$
Gaussian states in tripartite CV quantum information, and the
``uniqueness'' of their entanglement under manifold interpretations
as discussed in Sec.~\ref{secpromis}, much on the same footage as
the ``uniqueness'' of entanglement in symmetric (mixed) two-mode
Gaussian states (see Sec.~\ref{telebass})

\subsection*{Entanglement and optimal fidelity for nonsymmetric
Gaussian resources?}

Throughout this whole section, we have only dealt with completely
symmetric resource states, due to the invariance requirements of the
considered protocol. In Ref. \cite{telepoppate}, the question
whether expressions like  \eq{et}, connecting the optimal
teleportation fidelity to the entanglement in the shared resource,
were valid as well for nonsymmetric entangled resource states, was
left open (see also Ref. \cite{pirandolareview}). In
Sec.~\ref{sectlc}, devoted to telecloning, we will show with a
specific counterexample that this is {\em not} the case, not even in
the simplest case of $N=2$.

\section{$\mathbf{1 \boldsymbol\rightarrow 2}$ telecloning with bisymmetric and
nonsymmetric resources}\label{sectlc}

Quantum {\em telecloning} \cite{telequb} among $N+1$ parties is
defined as a process in which one party (Alice) owns an unknown
quantum state, and wants to distribute her state, via teleportation,
to all the other $N$ remote parties. The no-cloning theorem
\cite{nocloning1,nocloning2} yields that the $N$ remote clones can
resemble the original input state only with a finite, nonmaximal
fidelity. In CV systems, $1 \rightarrow N$ telecloning of arbitrary
coherent states was proposed in Ref. \cite{telecloning}, involving a
special class of $(N+1)$-mode multiparty entangled Gaussian states
(known as ``multiuser quantum channels'') shared as resources among
the $N+1$ users. The telecloning is then realized by a succession of
standard two-party teleportations between the sender Alice and each
of the $N$ remote receivers, exploiting each time the corresponding
reduced two-mode state shared by the selected pair of parties.

Depending on the symmetries of the shared resource, the telecloning
can be realized with equal fidelities for all receivers ({\em
symmetric} telecloning) or with unbalanced fidelities among the
different receivers ({\em asymmetric} telecloning). In particular,
in the first case, the needed resource must have complete invariance
under mode permutations in the $N$-mode block distributed among the
receivers: the resource state has to be thus a $1 \times N$
bisymmetric state \cite{adescaling,unitarily} (see Fig.
\ref{figbasset}).

In this manuscript we specialize on $1 \rightarrow 2$ telecloning,
where Alice, Bob and Claire share a tripartite entangled three-mode
Gaussian state and Alice wants to teleport arbitrary coherent states
to Bob and Claire with certain fidelities. As the process itself
suggests, the crucial resource enabling telecloning is not the
genuine tripartite entanglement (needed instead for a successful
`multidirectional' teleportation network, as shown in the previous section), but the
couplewise entanglement between the pair of modes $1|2$ and $1|3$
[if the sender (Alice) owns mode $1$, while the receivers (Bob and Claire) own
modes $2$ and $3$].

Let us notice that, very recently, the first experimental
demonstration of unconditional symmetric $1 \rightarrow 2$
telecloning of unknown coherent states has been achieved by
Furusawa's group \cite{exptelecloning}, with a fidelity for each
clone of ${\cal F} = 0.58 \pm 0.01$, surpassing the classical
threshold of $0.5$, \eq{fcl}. This experimental milestone has raised
renewed interest towards CV quantum communication \cite{press}.
Moreover, in keep with the general spirit of the paper, the context
of CV telecloning constitutes here the proper testground to
investigate the operational significance of the proposed
entanglement measures. The present analysis will lead to
conclusively show that the correspondence between entanglement and
teleportation fidelity does not extend to non-symmetric Gaussian
states.

\subsection{Symmetric telecloning}\label{telebass}

Let us first analyze the case of symmetric telecloning, occurring when Alice aims at sending two copies
of the original state with equal fidelities to Bob and Claire.
In this case it has been
proven \cite{cerfclon,iblisdir,telecloning} that Alice can teleport
an arbitrary coherent state to the two distant twins Bob and Claire
(employing a Gaussian cloning machine)
 with the maximal fidelity
\begin{equation}\label{f23}
{\cal F}_{\max}^{1 \rightarrow 2}=\frac{2}{3}\,.
\end{equation}
This argument inspired the introduction of the `no-cloning
threshold' for two-party teleportation \cite{grosgran}, basically
stating that only a fidelity greater than $2/3$ (thus greater than
the previously introduced threshold of $1/2$, which implies the
presence of entanglement) ensures the realization of actual
two-party quantum teleportation of a coherent state. In fact, if the
fidelity falls in the range $1/2 < {\cal F} < 2/3$, then Alice could
have kept a better copy of the input state for herself, or sent it
to a `malicious' Claire. In this latter case, the whole process
would result into an asymmetric telecloning, with a fidelity ${\cal
F} > 2/3$ for the copy received by Claire. It is worth remarking
that, as already mentioned, two-party CV teleportation beyond the
no-cloning threshold has been also recently demonstrated
experimentally, with a fidelity ${\cal F} = 0.70 \pm 0.02$
\cite{furunew}. Another important and surprising remark is that the
fidelity of $1 \rightarrow 2$ cloning of coherent states, given by
\eq{f23}, is {\em not} the optimal one. As recently shown in Ref.
\cite{clingon}, using non-Gaussian operations as well, two identical
copies of an arbitrary coherent state can be obtained with optimal
single-clone fidelity ${\cal F} \approx 0.6826$.

In our setting, dealing with Gaussian states and Gaussian operations
only, \eq{f23} represents the maximum achievable success for
symmetric $1 \rightarrow 2$ telecloning of coherent states. As
previously anticipated,  the {\em basset hound states} $\sig_B^p$ of
Sec. \ref{secbas} are  the best suited resource states for this
task. Such states belong to the family of multiuser quantum channels
introduced in Ref. \cite{telecloning}, and are $1 \times 2$
bisymmetric pure states, parametrized by the single-mode mixedness
$a$ of mode $1$. In particular, it is interesting to study how the
single-clone telecloning fidelity behaves compared with the actual
amount of entanglement in the $1|l$ ($l=2,3$) nonsymmetric two-mode
reductions of $\sig_B^p$ states.

A brief excursus has to be made here. Setting, as usual, all first
moments to zero, the fidelity of two-user teleportation of arbitrary
single-mode Gaussian states exploiting two-mode Gaussian resources
can be computed directly from the respective CMs \cite{fiuratele}.
Being $\sig_{in}$ the CM of the unknown input state, and
\begin{equation}\label{sig2}
\sig_{ab} = \left(
              \begin{array}{cc}
                \sig_a & \eps_{ab} \\
                \eps_{ab}^{\sf T} & \sig_b \\
              \end{array}
            \right)\,,
\end{equation}
the CM of the shared two-mode resource, and defining the matrix
$\gr{\xi} = {\rm diag}\{-1\,,1\}$, the fidelity reads
\cite{fiuratele}
\begin{equation}\label{ficm}
{\cal F} = \frac2{\sqrt{\det \gr\Sigma}}\,,\qquad \gr\Sigma \equiv
2\, \sig_{in} + \gr{\xi} \sig_a \gr{\xi} + \sig_b + \gr{\xi}
\eps_{ab} + \eps_{ab}^{\sf T} \gr{\xi}\,.
\end{equation}

\begin{figure}[t!]
\centering{
\includegraphics[width=10cm]{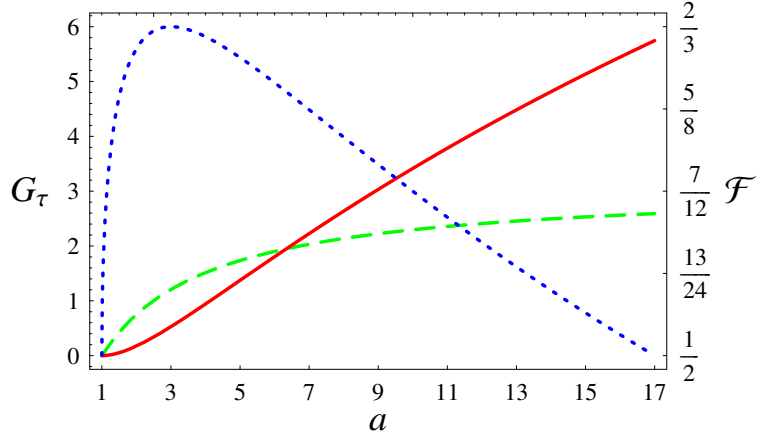}
\caption{Bipartite entanglement $G_\tau^{1|l}$ (dashed line) in
$1|l$ ($l=2,3$) two-mode reductions of basset hound states, and
genuine tripartite entanglement $G_\tau^{res}$ (solid line) among
the three modes, versus the local mixedness $a$ of mode $1$.
Entanglements are quantified by the Gaussian contangle. The fidelity
${\cal F}^{1 \rightarrow 2}_{sym}$ of symmetric $1 \rightarrow 2$
telecloning employing basset hound resource states is plotted as
well (dotted line, scaled on the right axis), reaching its optimal
value of $2/3$ for $a=3$.} \label{fitette}}
\end{figure}

In our case, $\sig_{in}= \id_2$ because Alice is teleporting
coherent states, while the resource $\sig_{ab}$ is obtained
discarding either the third ($a=1$, $b=2$) or the second ($a=1$,
$b=3$) mode from the CM $\sig_B^p$ of basset hound states. From
Eqs.~(\ref{bassigl}, \ref{basseps}, \ref{ficm}), the single-clone
fidelity for symmetric $1 \rightarrow 2$ telecloning exploiting
basset hound states is:
\begin{equation}\label{fitsym}
{\cal F}^{1 \rightarrow 2}_{sym} = \frac{4}{3 a - 2 \sqrt{2}
\sqrt{a^2 - 1} + 5}\,.
\end{equation}
Notice, remembering that each of modes $2$ and $3$ contains an
average number of photons $\bar n = (a-1)/2$, that \eq{fitsym} is
the same as Eq.~(19) of Ref. \cite{paris}, where a production scheme
for three-mode Gaussian states by interlinked nonlinear interactions
in $\chi^{(2)}$ media is presented, and the usefulness of the
produced resources for $1 \rightarrow 2$ telecloning is discussed as
well. The basset hound states realize an optimal symmetric cloning
machine, {\ie}the fidelity of both clones saturates \eq{f23}, for
the finite value $a=3$. Surprisingly, with increasing $a > 3$, the
fidelity \eq{fitsym} starts decreasing, even if the two-mode
entanglements \eq{g11bh} in the reduced (nonsymmetric) bipartitions
of modes $1|2$ and $1|3$, as well as the genuine tripartite
entanglement \eq{gtauresbh}, increase with increasing $a$. As shown
in Fig. \ref{fitette}, the telecloning fidelity is not a monotonic
function of the employed bipartite entanglement. Rather, it roughly
follows the difference $G_\tau^{1|l}-G_\tau^{res}$, being maximized
where the bipartite entanglement is stronger than the tripartite
one.  This fact heuristically confirms that in basset hound states
bipartite and tripartite entanglements are competitors, meaning that
the CV entanglement sharing in these states is not promiscuous, as
described in Sec. \ref{secbas}.

The example of basset hound states represents a clear hint that the
teleportation fidelity with general two-mode (pure or mixed)
nonsymmetric resources is {\em not} monotone with the entanglement.
Even if an hypothetical optimization of the fidelity over the local
unitary operations could be performed (on the guidelines of
\cite{telepoppate}), it would entail a fidelity growing up to $2/3$
and then staying constant while entanglement increases, which means
that no direct estimation of the entanglement can be extracted from
the nonsymmetric teleportation fidelity, at variance with the
symmetric case (see the previous section). More precisely, it can be
easily shown that the direct relationship between negativity and
teleportation fidelity valid for symmetric states cannot carry over
to nonsymmetric resources as well. In point of fact, applying it to
the $1|l$ ($l=2,3$) two-mode reduced resources obtained from basset
hound states, would imply an ``optimal'' fidelity reaching $3/4$ in
the limit $a \rightarrow \infty$. But this value is impossible to
achieve, even considering non-Gaussian cloning machines
\cite{clingon}: thus, the simple relation between teleportation
fidelity and entanglement, formalized by \eq{et}, fails to hold for
nonsymmetric resources, even in the basic two-mode instance. Let us
mention that the relationship between entanglement and teleportation
fidelity had already been partially addressed in Ref.~\cite{kimlee}
(which also points out other counterintuitive features occurring for
asymmetric mixed resources, like the enhancement of fidelity under
suitable noisy pre-processing). However, Ref.~\cite{kimlee} actually
addresses the relationship between teleportation fidelity and the
squeezing parameter of an asymmetrically decohered two-mode squeezed
state (used as the entangled resource). We remark that such a
quantity, while being related to the entanglement of the state is
not, by itself, a proper entanglement quantifier ({\em e.g.}, it
does not determine the negativity of the nonsymmetric state).

This somewhat controversial result can be to some extent interpreted
as follows. For symmetric Gaussian states, there exists a `unique
type' of bipartite CV entanglement. In fact, measures such as the
logarithmic negativity (quantifying the violation of the
mathematical PPT criterion), the entanglement of formation (related
to the entanglement cost, and thus quantifying how expensive is the
process of creating a mixed entangled state through LOCC), and the
degree of EPR correlation (quantifying the correlations between the
entangled degrees of freedom) are {\em all} completely equivalent
for such states, being monotonic functions of only the smallest
symplectic eigenvalue $\tilde{\nu}_-$ of the partially transposed
CM. As we have seen, this equivalence extends also to the efficiency
of two-user quantum teleportation, quantified by the fidelity
optimized over local unitaries. For nonsymmetric states, the chain
of equivalences breaks down. In hindsight, this could have been
somehow expected, as there exist several inequivalent but legitimate
measures of entanglement, each of them capturing distinct aspects of
the quantum correlations (one could think, for instance, of the
operative difference existing between the definitions of distillable
entanglement and entanglement cost). In the specific instance of
nonsymmetric two-mode Gaussian states, it has been shown that the
negativity is neither equivalent  to the (Gaussian) entanglement of
formation (the two measures may induce inverted orderings on this
subset of entangled states) \cite{ordering}, nor to the EPR
correlation \cite{extremal}. It is thus justified that a process
like teleportation emphasizes a distinct aspect of the entanglement
encoded in nonsymmetric resources. Notice also that the richer and
more complex entanglement structure of non symmetric states, as
compared to that of symmetric states, reflects a crucial operational
difference in the respective (asymmetric and symmetric)
teleportation protocols. While in the symmetric protocols the choice
of sender and receiver obviously does not affect the fidelity, this
is no longer the case in the asymmetric instance: this physical
asymmetry between sender and receiver properly exemplifies the more
complex nature of the two-mode asymmetric entanglement.

\subsection{Asymmetric telecloning}

\begin{figure}[t!]
\centering{\includegraphics[width=9cm]{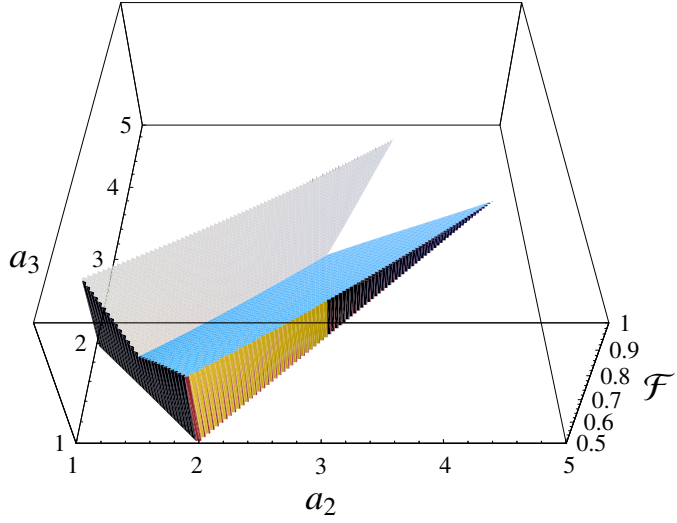}%
\caption{\label{fiaclo}Fidelities for asymmetric telecloning with
three-mode pure Gaussian resources, at a fixed $a_1=2$, as functions
of $a_2$ and $a_3$, varying in the allowed range of parameters
constrained by \ineq{triangleprim} (see also Fig. \ref{pompilio}).
The darker surface on the right-hand side of the diagonal $a_2=a_3$
(along which the two surfaces intersect) is the fidelity of Bob's
clone, ${\cal F}_{asym:2}^{1 \rightarrow 2}$, while the lighter,
`mirror-reflected' surface on the left-hand side of the diagonal is
the fidelity of Claire's clone, ${\cal F}_{asym:3}^{1 \rightarrow
2}$. Only nonclassical fidelities ({\ie}${\cal F} > 1/2$) are
shown.}}
\end{figure}

In this section we focus on the {\em asymmetric} telecloning of
coherent states, through generic pure three-mode Gaussian states
shared as resources among the three parties. Considering states in
standard form (see Sec.~\ref{secallot} and Ref.~\cite{3mpra}),
parametrized by the local single-mode mixednesses $a_i$ of modes
$i=1,2,3$, the fidelity ${\cal F}_{asym:2}^{1 \rightarrow 2}$ of
Bob's clone (employing the $1|2$ two-mode reduced resource) can be
computed from \eq{ficm} and reads
\begin{eqnarray}\label{fiasym2}
\hspace*{-2cm} {\cal F}_{asym:2}^{1 \rightarrow 2} &=& 2\ \Bigg\{-2
a_3^2 + 2 a_1 a_2 + 4 \left(a_1 + a_2\right) +
          3 \left(a_1^2 +
                a_2^2\right) \\
 \hspace*{-2cm}  &-& \left(a_1 + a_2 +
                  2\right) \sqrt{\frac{[(a_1 + a_2 - a_3)^2 -
                          1] [(a_1 + a_2 + a_3)^2 -
                        1]}{a_1 a_2}} + 2
                        \Bigg\}^{-\frac{1}{2}}\,,
                        \nonumber
\end{eqnarray}
Similarly, the fidelity ${\cal F}_{asym:3}^{1 \rightarrow 2}$ of
Claire's clone can be obtained from \eq{fiasym2} by exchanging the
roles of ``$2$'' and ``$3$''.

\begin{figure}[t!]
\centering{\includegraphics[width=9cm]{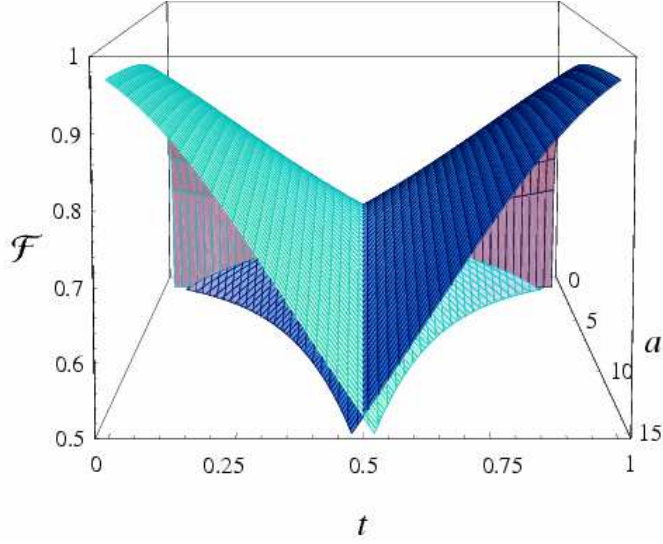}%
\caption{\label{fiapeto}Optimal fidelities for asymmetric
telecloning with three-mode pure Gaussian resources, as functions of
the single-mode mixedness $a$ of mode $1$, and of the parameter $t$
determining the local mixednesses of the other modes, through
Eqs.~(\ref{asybest},\ref{basy}). The darker, rightmost surface is
the optimal fidelity of Bob's clone, ${\cal F}_{asym:2}^{opt:1
\rightarrow 2}$, while the lighter, leftmost surface is the optimal
fidelity of Claire's clone, ${\cal F}_{asym:3}^{opt:1 \rightarrow
2}$. Along the intersection line $t=1/2$ the telecloning is
symmetric. Only nonclassical fidelities ({\ie}${\cal F}
> 1/2$) are shown.}}
\end{figure}

It is interesting to explore the space of parameters
$\{a_1,\,a_2,\,a_3\}$ in order to find out which three-mode states
allow for an asymmetric telecloning with the fidelity of one clone
above the symmetric threshold of $2/3$, while keeping the fidelity
of the other clone above the classical threshold of $1/2$. Let us
keep $a_1$ fixed. With increasing difference between $a_2$ and
$a_3$, one of the two telecloning fidelities increases at the
detriment of the other, while with increasing sum $a_2+a_3$ both
fidelities decrease to fall eventually below the classical
threshold, as shown in Fig. \ref{fiaclo}. The asymmetric telecloning
is thus {\em optimal} when the sum of the two local mixednesses of
modes $2$ and $3$ saturates its lower bound. From
\ineq{triangleprim}, the optimal resources must have
\begin{equation}\label{asybest}
a_3 = a_1 - a_2 + 1\,,
\end{equation}
A suitable
parametrization of these states is obtained setting $a_1 \equiv a$
and
\begin{equation}\label{basy}
a_2 = 1+(a-1)t\,,\qquad 0 \le t \le 1\,.
\end{equation}
For $t < 1/2$ the fidelity of Bob's clone is smaller than that of
Claire's one, ${\cal F}_{asym:2}^{1 \rightarrow 2} <  {\cal
F}_{asym:3}^{1 \rightarrow 2}$, while for $t > 1/2$ the situation is
reversed. In all the subsequent discussion, notice that Bob and
Claire swap their roles if $t$ is replaced by $1-t$. For $t=1/2$,
the asymmetric resources reduce to the bisymmetric basset hound
states useful for symmetric telecloning. The optimal telecloning
fidelities then read
\begin{equation}\label{fiazz12}
\hspace*{-1cm}
\begin{array}{c} {\cal F}_{asym:2}^{opt:1 \rightarrow 2} =\frac{2}{\sqrt{(a + 3)^2 +
(a - 1)^2 t^2 + 2 (a - 1) (3 a + 5) t-4 \sqrt{\left(a^2 - 1\right)
t} [a + (a - 1) t + 3]}}\,,
  \end{array}
\end{equation}
and similarly for ${\cal F}_{asym:3}^{opt:1 \rightarrow 2}$
replacing $t$ by $1-t$. The two optimal fidelities are plotted in
Fig. \ref{fiapeto}

With these pure nonsymmetric resources, further optimizations can be
performed depending on the needed task. For instance, one may need
to implement telecloning with the highest possible fidelity of one
clone, while keeping the other nonclassical. This problem is of
straightforward solution, and yields optimal asymmetric resources
with
\begin{equation}\label{atas}
a=\frac{7}{2}\,,\quad t=\frac{4}{5}\,.
\end{equation}
In this case the fidelity of Claire's clone saturates the classical
threshold, ${\cal F}_{asym:3}^{opt:1 \rightarrow 2}=1/2$, while the
fidelity of Bob's clone reaches ${\cal F}_{asym:3}^{opt:1
\rightarrow 2}=4/5$, which is the maximum allowed value for this
setting \cite{fiuraclon}. Also, choosing $t=1/5$, Bob's fidelity
gets classical and Claire's fidelity is maximal.

In general, a telecloning with ${\cal F}_{asym:2}^{opt:1 \rightarrow
2} \ge 2/3$ and ${\cal F}_{asym:3}^{opt:1 \rightarrow 2} \ge 1/2$ is
possible only in the window
\begin{equation}\label{awindow}
 1.26 \approx 2 \sqrt{2} \left[2 - \sqrt{1 + \sqrt{2}}\right]  \le a
\le 2 \sqrt{2} \left[2 + \sqrt{1 + \sqrt{2}}\right] \approx 10.05\,
\end{equation}
and, for each $a$ falling in the region defined by \ineq{awindow},
in the specific range
\begin{equation}\label{trange}
\frac{a - 2 \sqrt{a + 1} + 2}{a - 1} \le t \le \frac{2
\left(\sqrt{2} \sqrt{a + 1} - 2\right)}{a - 1}\,.
\end{equation}
For instance, for $a=3$, the optimal asymmetric telecloning (with
Bob's fidelity above no-cloning and Claire's fidelity above
classical bound) is possible in the whole range $1/2 \le t \le
2\sqrt{2}-1$, where the boundary $t=1/2$ denotes the basset hound
state realizing optimal symmetric telecloning (see Fig.
\ref{fitette}). The sum ${\cal S}^{opt:1 \rightarrow 2}={\cal
F}_{asym:2}^{opt:1 \rightarrow 2} + {\cal F}_{asym:3}^{opt:1
\rightarrow 2}$ can be maximized as well, and the optimization is
realized by values of $a$ falling in the range $2.36 \lesssim a \le
3$, depending on $t$. The absolute maximum of ${\cal S}^{opt:1
\rightarrow 2}$ is reached, as expected, in the fully symmetric
instance $t=1/2$, $a=3$,
and yields ${\cal S}^{opt:1 \rightarrow 2}_{\max} = 
4/3$.

We finally recall that optimal three-mode Gaussian
resources, can be produced by implementing the allotment operator
(see Sec. \ref{secallot}), and employed to perform all-optical
symmetric and asymmetric telecloning machines
\cite{telecloning,fiuraclon}.

\section{Conclusions}\label{conclu}
In the present paper and in Ref. \cite{3mpra},
we have aimed at providing a (to a good extent)
comprehensive treatment on the
characterization, quantification and experimental generation of
genuine multipartite entanglement in three-mode Gaussian states of
CV systems, including relevant, practical
implementations in the context of CV quantum information.

Similarly with what happens for bipartite entanglement in symmetric
Gaussian states, but at striking variance with the discrete-variable
scenario (in particular for systems of three qubits), we have shown
that there is a {\em unique} kind of genuine tripartite entanglement
in pure, symmetric, three-mode Gaussian states, which combines a
mathematical significance in the context of entanglement sharing
with an operational interpretation in terms of teleportation
experiments. This fact is a remarkable consequence of the
restriction to Gaussian states, as the proven existence, for
infinite dimensional systems, of infinitely many mutually
stochastic-LOCC-incomparable states (even under the bounded energy
and finite information exchange condition) suggests \cite{sloccv}.
Even more strikingly, this tripartite entanglement distributes in a
{\em promiscuous} way, being enhanced by the presence of  bipartite
entanglement in any two-mode reduction. Here we have shown that the
promiscuity of CV entanglement survives even for non-pure states
like noisy GHZ/$W$ states, with purities down to $0.2$. However,
with increasing mixedness the structure of three-party entanglement
enriches, as tripartite bound entangled states can exist even in the
fully symmetric instance. For nonsymmetric (pure or mixed)
three-mode states, the promiscuity fades leaving room for a more
traditional entanglement sharing structure. We further remark that
for all fully inseparable Gaussian states (which represent one of
the five possible separability classes \cite{kraus}), the residual
Gaussian contangle, or \emph{arravogliament}, is a {\em bona fide},
computable measure of genuine tripartite entanglement, as first
demonstrated in Refs. \cite{contangle,3mpra}, and evaluated in
several explicit instances here.

The core of this paper has been devoted to providing several
examples of tripartite entangled Gaussian states, relevant for their
entanglement properties and/or for applications in CV quantum
information. For each of them, we have computed the tripartite
entanglement explicitly, and dedicated great attention to quantum
state engineering, proposing schemes for the experimental production
of the considered states. In particular we developed a
 self-contained procedure to engineer (up to local operations) arbitrary pure
three-mode Gaussian states by means of linear optics, based on the
distribution, or {\em allotment} of two-mode entanglement among
three modes.

In the last part of this work we have investigated the
potentialities of the introduced families of three-mode Gaussian
states for the implementation of multipartite quantum communication
protocols with continuous variables. This has allowed us to focus on
the physical significance of the applied `genuine multipartite'
entanglement measure (arravogliament), shifting the analysis's testing
ground from the domain of mathematical convenience to that of
operational effectiveness.

\addcontentsline{toc}{section}{Acknowledgments}
\section*{Acknowledgments}

Financial support from MIUR, INFN, and CNR is acknowledged. G A is
grateful to Samanta Piano for her encouraging advice, and for
reminding him of the sentence quoted in the caption of
Fig.~\ref{figbasset}. A S is currently a Marie Curie fellow at
Imperial College London; he also acknowledges EPSRC for financial
support (through the QIP-IRC) and the Centre for Mathematical
Sciences of the University of Cambridge for kind hospitality.

\end{document}